\documentclass[11pt,a4paper,oneside]{article}
\usepackage[utf8]{inputenc}
\usepackage[a4paper,textwidth=16cm,textheight=23cm,centering]{geometry}
\usepackage[affil-it]{authblk}

\usepackage{appendix}
\usepackage{todonotes}
\usepackage{amsmath}
  \usepackage{comment}
  \numberwithin{equation}{section}

\usepackage[]{amssymb}
\usepackage{amsfonts}
\usepackage{mathrsfs}
\usepackage{amsthm}

\theoremstyle{definition}

	\newtheorem{rmk}[equation]{Remark}

\DeclareMathOperator{\N}{\mathbb{N}}
\DeclareMathOperator{\Z}{\mathbb{Z}}

\DeclareMathOperator{\ver}{\mathrm{Vert}}
\DeclareMathOperator{\edge}{\mathrm{Edge}}

\newcommand{\mN}{\mathcal{N}}

\newcommand{\mI}{\mathcal{I}}

\newcommand{\dd}{\mathrm{d}}
\newcommand{\ii}{\mathtt{i}}

\newcommand{\tr}{\mathrm{Tr}}
\newcommand{\sli}{\dot{\iota}}

\def\Xint#1{\mathchoice
   {\XXint\displaystyle\textstyle{#1}}%
   {\XXint\textstyle\scriptstyle{#1}}%
   {\XXint\scriptstyle\scriptscriptstyle{#1}}%
   {\XXint\scriptscriptstyle\scriptscriptstyle{#1}}%
   \!\int}
\def\XXint#1#2#3{{\setbox0=\hbox{$#1{#2#3}{\int}$}
     \vcenter{\hbox{$#2#3$}}\kern-.52\wd0}}
\def\dashint{\Xint-}

\usepackage[english]{babel}
\usepackage{caption}
\usepackage{enumerate}

\usepackage{graphicx}
\usepackage{xcolor}
\usepackage[
  colorlinks=true,
  citecolor=violet,
  linkcolor=blue,
  urlcolor=blue
]{hyperref}

\usepackage{tikz}
\usetikzlibrary{decorations.pathreplacing}
\usetikzlibrary{er,positioning,calc,shapes.geometric,decorations.markings}

\tikzset{u/.style={circle, draw=black, fill=white,inner sep=2 pt, minimum size=2 pt,},f/.style={square, draw=black, fill=white,ultra thick,inner sep=4 pt, minimum size=2 pt,}}
\tikzset{uf/.style={regular polygon,regular polygon sides=4, draw=black, fill=white,inner sep=3 pt, minimum size=4 pt,}}
\newcommand{\loopedge}[3]{\draw (#1) to[out=#2,in=#3,looseness=7] (#1);}

\DeclareMathOperator{\HS}{HS}

\DeclareMathOperator{\PE}{PE}
\newcommand\aafootnote[1]{%
  \begingroup
  \renewcommand\thefootnote{}\footnote{#1}%
  \addtocounter{footnote}{-1}%
  \endgroup
}
\usepackage[numbers,compress,square,comma]{natbib}
\begin{document}
\bibliographystyle{mystyle}
\captionsetup[figure]{labelfont={sc,small},labelformat={default},labelsep=period,font=small}

{\pagenumbering{roman} 
\renewcommand*{\thefootnote}{\fnsymbol{footnote}}
	
\title{\huge\textbf{Superconformal index of large $N$ and long quivers}}

\author[$\spadesuit$]{Mohammad Akhond}
\affil[$\spadesuit$]{\small \it Sezione INFN Roma “Tor Vergata” \& Dipartimento di Fisica,\protect\\ \small\it Universit\`a di Roma “Tor Vergata”, Via della Ricerca Scientifica 1, 00133, Roma, Italy}

\author[$\clubsuit$]{Leonardo Santilli}
\affil[$\clubsuit$]{\small\it Center for Mathematics and Interdisciplinary Sciences,\protect\\ \small\it Fudan University, Shanghai, 200433, China}	
\affil[$\clubsuit$]{\small\it Shanghai Institute for Mathematics and Interdisciplinary Sciences (SIMIS),\protect\\ \small\it Shanghai, 200433, China}	

\date{ \hspace{8pt} }
	
	\maketitle
	\thispagestyle{empty}
\aafootnote{$^{\spadesuit}$\texttt{akhond@roma2.infn.it}}
\aafootnote{$^{\clubsuit}$\texttt{santilli at simis dot cn}}
			
\begin{abstract}
We develop a universal framework for computing generating functions in quiver gauge theories in the large $N$ limit. We apply this method to the superconformal index of 4d $\mathcal{N}=2$ gauge theories, as well as to the 0-instanton index of 5d $\mathcal{N}=1$ theories, both for finite quivers of arbitrary shape and in the long quiver limit, including both special unitary and orthosymplectic gauge groups, demonstrating its versatility. In the Hall--Littlewood limit, we show that the series expansion of the resulting index captures the expected Higgs branch chiral ring generators.
Our exact expressions in the long quiver limit provide new predictions for the supergravity index of the holographically dual warped AdS$_5$ and AdS$_6$ backgrounds.
\end{abstract}

	\clearpage
	{
	\tableofcontents}
\thispagestyle{plain}
}	
	\clearpage
	\pagenumbering{arabic}
	\setcounter{page}{1}
		\renewcommand*{\thefootnote}{\arabic{footnote}}
		\setcounter{footnote}{0}
\section{Introduction and summary}

Quantum field theories often admit a (semi)classical limit, when one of their parameters is taken to an extreme value. Familiar examples are the perturbative expansions at weak coupling, large number of colours \cite{tHooft:1973alw, tHooft:1974pnl, Brezin:1977sv}, or large number of flavours \cite{Veneziano:1976wm}. Recent advances in supersymmetric theories \cite{Uhlemann:2019ypp,Uhlemann:2019lge,Uhlemann:2020bek,Coccia:2020wtk,Coccia:2021lpp,Fatemiabhari:2022kpv,Akhond:2022awd,Akhond:2022oaf,Apruzzi:2022nax,Santilli:2023fuh,Nunez:2023loo,He:2024djr} have revealed a novel regime of quiver gauge theories in which both the rank $N$ of the gauge groups and the number $L$ of gauge nodes become large. This \emph{long quiver} limit exhibits remarkable simplifications and has already led to new exact results for sphere free energies of theories with 8 supercharges in dimensions three to five. A review of the recent developments is found in \cite{Santilli:2025zum}.

\paragraph{}Given the success of the long quiver expansion in the case of the sphere partition function, it is natural to extend these techniques to supersymmetric indices \cite{Romelsberger:2005eg,Kinney:2005ej,Bhattacharya:2008zy}. The main goal of this paper is to initiate the study of the long quiver limit of superconformal indices. The setting for this study is the index of 4d $\mathcal{N}=2$ theories, while the 0-instanton index of 5d $\mathcal{N}=1$ theories admits a similar solution. 

\paragraph{} From a purely mathematical perspective, the novelty of the long quiver index is that it is a unitary matrix model, whereas the sphere partition functions that have been analysed previously are Hermitian matrix models. We expect that the techniques developed here are of further applications to broader contexts where unitary quiver matrix models arise, for instance by considering theories with less, or no supersymmetry at all. With this in mind, our analysis begins with an abstract introduction of the class of matrix models of interest in section \ref{sec:QuiverMM}. The superconformal index is subsequently introduced, and shown to fit within this general class in section \ref{sec:SCIdata}.

\paragraph{} Along the way, we derive a universal formula for the large $N$ limit of these quiver matrix models, for any shape and size of the quiver and for any classical semisimple gauge group and matter content with up to two gauge indices. The supersymmetric indices of quiver gauge theories with 8 supercharges, not necessarily conformal, are obtained as specialisations of this result.

\paragraph{} The strongly coupled nature of the quiver gauge theories in question provides sufficient intrinsic value to the exact results obtained in the long quiver limit. Nevertheless, the reader should be informed of some wider applications which serve as background motivations. The 4d and 5d long quivers studied in this work have holographic dual warped AdS solutions in type IIA \cite{Gaiotto:2009gz, Reid-Edwards:2010vpm,Aharony:2012tz,Nunez:2019gbg} and type IIB string theory \cite{DHoker:2016ujz,DHoker:2017mds,DHoker:2017zwj, Legramandi:2021uds,Legramandi:2021aqv} respectively, and the long quiver limit corresponds to the weakly curved supergravity regime of the dual string theory. As such, the results of this paper provide new predictions for precision holography. 

\paragraph{} Additionally, long circular quivers are conjectured to deconstruct higher dimensional theories, notably the (2,0) theory \cite{Arkani-Hamed:2001wsh}. The long quiver limit of the index of circular $\mathcal{N}=2$ quivers was previously computed in the Hall--Littlewood limit and argued to be consistent with expectations from the (2,0) side \cite{Hayling:2017cva}. A refined expression of the superconformal index for long circular quivers is provided by a direct application of \eqref{eq:IlongQ}. For recent work on other aspects of long circular quivers see for instance \cite{Beccaria:2023qnu,Korchemsky:2025eyc} and references therein. 

\subsection*{Summary of main results}
In the remainder of this introduction, we highlight the main results of this paper, whose details are found in the bulk of the text.
\paragraph{Large N index for finite size quivers.} For quiver theories with special unitary gauge groups and arbitrary number of fundamental, bifundamental and adjoint hypermultiplets, we find 
\begin{equation}
    \left. \mI \right\rvert_{N\to\infty} =\PE\left[ \sli_{H}(\mathbf{p})^2\sum_{i,j} F_iF_j [D_1]_{ij}  \right] \prod_{n\geq 1}\det D_n\;,
\end{equation}
where $\PE[\cdot]$ denotes the plethystic exponential defined in \eqref{eq:plethystic exp}, $\sli_H(\mathbf{p})$ is the single-letter index of the hypermultiplet, the inner sum runs over the vertices of the quiver, $F_i$ are the ranks of the framing nodes, and the matrices $D_n$, which depend on the fugacities, contain the information about gauge groups and edges of the quiver. We refer to sections \ref{sec:IndexSUresult}-\ref{sec:IndexSOSpresult} for the most general expressions for special unitary and orthosymplectic quivers, respectively.

\paragraph{Index of large $N$ and long quivers.} For long linear or circular quivers, we find 
 \begin{equation}
\left.\mI \right\rvert_{\substack{N \to \infty \\ L\to \infty}} =\PE\left[ L^{-1}\;  \sli_{H}(\mathbf{p})  \sum_{i,j}F_iF_j G_1 \left( \frac{i}{L}, \frac{j}{L}\right)  \right] \prod_{n\geq 1}\left(\frac{t_n}{\sli_H (\mathbf{p}^n)}\right)^L \;, 
\end{equation}
where $G_1$ is the Green's function for a one-dimensional Poisson equation --- the saddle point equation for the eigenvalue densities --- given explicitly in section \ref{sec:exactILong}, and $t_n$ is defined in \eqref{eq:deftnHL}. The infinite product, which is the long quiver limit of the determinant, contributes to the same order in $L$ but is subleading in $N$, thus it should be omitted when $F_i = \mathcal{O}(N)$.
Conversely for long circular quivers \emph{without} framing, the plethystic exponential trivialises and the determinant \emph{does} contribute to leading order.

\paragraph{Structure of the paper.} Section \ref{sec:General} introduces the matrix model machinery, first in an abstract notation, which is then connected to the index. The large $N$ limit of these general matrix models is set up in section \ref{sec:LargeNMM}. Section \ref{sec:FiniteL} is dedicated to the index of large rank, but finite length quivers, with section \ref{sec:IndexSUresult} dedicated to special unitary theories and section \ref{sec:IndexSOSpresult} to orthosymplectic theories. The length of the quiver is subsequently promoted to be large in section \ref{sec:LargeL}. Applications deemed interesting in the authors' view are presented in the form of a number of examples in section \ref{sec:applications}, some of which provide consistency tests of our general results, while others provide novel predictions. 

\paragraph{Note added.} While this work was nearing completion, we became aware of related work by Yuanyuan Fang, Jing Feng, and Dan Xie \cite{friends}. 
We are grateful to these authors for coordinating the submission to arXiv.

\section{General unitary quiver matrix model}\label{sec:General}
This section sets up the matrix model associated with the superconformal index of a quiver gauge theory with 8 supercharges in four and five dimensions. The structure of the perturbative part of the index across these different spacetime dimensions is fairly universal.\footnote{Although in five dimensions, we only treat the perturbative part of the index, and instanton corrections are not taken into account.} The expressions provided in this section allow a simultaneous treatment in the subsequent sections.
 
\subsection{Quiver matrix models setup}\label{sec:QuiverMM}
A quiver gauge theory with 8 supercharges is specified by a framed graph, whose sets of vertices and (unoriented) edges we denote, respectively, by $\ver$ and $\edge$. We label a node, or vertex by an index $i\in\ver$, and an edge by $e \in \edge$;
we will also write $\{ e : i\to j\} \subset \edge $ for the collection of edges connecting two nodes $i,j\in\ver$. Associated to each vertex, is a simple gauge group factor $SU(N_i)$, so that the total gauge group is
\begin{equation}\label{eq:Ggauge}
    G=\prod_{i\in\ver} SU(N_i)\;.
\end{equation}
Finally, each node carries an additional framing, labelled by an integer $F_i$. We will come back to the physical meaning of this data momentarily in section \ref{sec:SCIdata}.
\paragraph{}For ease of presentation, we first consider the group 
\begin{equation}\label{eq:GUgauge}
    G^{\prime}=\prod_{i\in\ver} U(N_i)\;.
\end{equation}
instead of \eqref{eq:Ggauge}; we will show at the end that (i) the large $N$ solution for $G^{\prime}$ is also a solution for $G$, and (ii) subtracting an overall contribution from the diagonal $U(1)$ at each vertex passes from $G^{\prime}$ to $G$.

\paragraph{} We define the quiver matrix model
\begin{equation}\label{eq:ZPExp}
    \mI = \int_{G^{\prime}} \dd\mu_{G^{\prime}}(X) f (X)
\end{equation}
with $\dd\mu_{G^{\prime}}$ the normalised Haar measure, which we express in terms of the eigenvalues as
\begin{equation}\label{eq:Haar}
    \dd\mu_{G^{\prime}}(X) = \frac{1}{C} \left[ \dd X \right] , \qquad \left[ \dd X \right] = \prod_{j\in\ver}\prod_{c=1}^{N_j} \frac{\dd x_{c,j}}{2\pi i x_{c,j}}\prod_{i\in \ver} \prod_{1 \le a<b \le N_i}\left\lvert x_{a,i}-x_{b,i}\right\rvert^2  ,
\end{equation}
with normalisation constant $C= \prod_{j\in\ver}\prod_{c=1}^{N_j} \oint [\dd X]$.
In \eqref{eq:ZPExp}, $f$ is a class function of the general form
\begin{equation}
\begin{aligned}\label{eq:fpreMM}
    f(X) = \exp &\left\{  \sum_{i\in\ver} \sum_{n=1}^{\infty} \left[\sum_{a=1}^{N_i}\frac{v^{i}_n}{n}\left( x^n_{a,i}+ x^{-n}_{a,i}\right) + \sum_{a=1}^{N_i}\sum_{b=1}^{N_i}\frac{h^{V,i}_n}{n}\left( \frac{x^n_{a,i}}{x^n_{b,i}}+ \frac{x^n_{b,i}}{x^n_{a,i}}\right)\right] \right.\\
    & \quad \left. + \sum_{\substack{ e\in \edge \\ e: i \to j }}\sum_{a=1}^{N_i}\sum_{b=1}^{N_j}\sum_{n=1}^{\infty}\frac{h^{e}_n}{n}\left( \frac{x^n_{a,i}}{x^n_{b,j}}+ \frac{x^n_{b,j}}{x^n_{a,i}}\right) \right\}\,.
\end{aligned}
\end{equation}
The Fourier coefficients $h^{e}_n$ include the contributions from edges $e:i \to j$, while $h^{V,i}_n$ contain the self-interactions of the vertices. We defer a more detailed explanation to section \ref{sec:SCIdata}. It is convenient to collect the coefficients of the double-trace terms by defining 
\begin{equation}\label{eq:hijgeneric}
    h^{ij}_n = \delta^{ij} h^{V,i}_n + \sum_{\{ e: i \to j \}} h_n^{e} ,
\end{equation}
so that the Fourier coefficients $h^{ij}_n$ include the contributions from all edges $\{e:i \to j\}$, plus the vertex self-interaction contributions when $i=j$. In particular 
\begin{align}
    h^{ij}_n &=h^{ji}_n , \label{eq:hsymmetric}
\end{align} 
and $h^{ij}_n=0$ if and only if $i \ne j$ and $\{e:i \to j\}=\emptyset$. Then, \eqref{eq:fpreMM} is written as
\begin{equation}\label{eq:fMM}
    f(X) = \exp \left[  \sum_{i\in\ver} \sum_{a=1}^{N_i}\sum_{n=1}^{\infty}\frac{v^{i}_n}{n}\left( x^n_{a,i}+ x^{-n}_{a,i}\right) + \sum_{\substack{ (i,j)\in \ver^2 \\ i \le j }}\sum_{a=1}^{N_i}\sum_{b=1}^{N_j}\sum_{n=1}^{\infty}\frac{h^{ij}_n}{n}\left( \frac{x^n_{a,i}}{x^n_{b,j}}+ \frac{x^n_{b,j}}{x^n_{a,i}}\right)\right] \,.
\end{equation}
The summation in the second term of \eqref{eq:fMM} is restricted to $i \le j$, in an arbitrary ordering of the vertex set, to avoid over-counting the edge contributions.\par
Parametrising the maximal torus of $G$ by $x_{a,i}=\exp\left( \ii \theta_{a,i}\right)$ and manipulating \eqref{eq:Haar}, we get:
\begin{equation}\label{eq:Zeff}
    \mI = \frac{1}{C}\prod_{j\in\ver}\exp \left( N_j \sum_{n=1}^{\infty} \frac{1}{n}\right) \prod_{c=1}^{N_j} \dashint_{-\pi}^{\pi} \frac{\dd \theta_{c,j}}{2\pi} \exp \left( S_{\mathrm{eff}} \right) ,
\end{equation}
where 
\begin{equation}\label{eq:SeffGeneral}
\begin{aligned}
S_{\mathrm{eff}} = \sum_{n =1}^{\infty}\frac{1}{n} &\left[ \sum_{i\in\ver} \sum_{a=1}^{N_i}\sum_{b=1}^{N_i} \left( h^{ii}_n -1 \right) \cos (n(\theta_{a,i}-\theta_{b,i})) \right. \\
& \left. + \sum_{\substack{(i,j)\in\ver^2 \\ j \ne i} } \sum_{a=1}^{N_i}\sum_{b=1}^{N_j} h^{ij}_n \cos (n(\theta_{a,i}-\theta_{b,j})) + 2 \sum_{i\in \ver} \sum_{a=1}^{N_i} v^{i}_n \cos (n\theta_{a,i}) \right].
\end{aligned}
\end{equation}
The symbol $\dashint_{-\pi}^{\pi}$ in \eqref{eq:Zeff} means that the hyperplanes $\theta_{a,i}=\theta_{b,i}$ for all $1 \le a<b\le N_i$ (for each $i$) have been removed from the integration domain. In the second line of \eqref{eq:SeffGeneral} we have used \eqref{eq:hsymmetric} to replace $2\sum_{i<j} (\cdots) $ by $\sum_{j \ne i} (\cdots)$.

\subsection{Superconformal index}\label{sec:SCIdata}

We now explain how the superconformal indices of 4d $\mN=2$ theories, and the perturbative indices of 5d $\mN=1$ theories, are recovered as particular cases of the matrix models defined in section \ref{sec:QuiverMM}. Here we simply state the correspondence, and refer to \cite{Rastelli:2016tbz,Gadde:2020yah} for reviews. Notice that we will not use the conformal symmetry at any point, hence our derivation applies to the supersymmetric index of any quiver gauge theory.\par
Consider a gauge theory with gauge group \eqref{eq:Ggauge}. $\forall i \in \ver$ we allow $F_i$ fundamental hypermultiplets of $SU(N_i)$, represented by an edge to the framing node; in addition we allow $g_i$ adjoint, $\Lambda_i$ rank-2 antisymmetric and $S_i$ rank-2 symmetric hypermultiplets charged under $SU(N_i)$, all represented by edges $e:i \to i$ of the graph. Moreover, edges $e:i \to j$ with $j \ne i$ correspond to hypermultiplets in the bifundamental representation of $SU(N_i)\times SU(N_j)$.\par
To reduce clutter, we focus on the case with fundamental, bifundamental and adjoint hypermultiplets, and comment on how to include hypermultiplets in the rank-2 symmetric and antisymmetric representations in Remarks \ref{rmk:rank2UN_1}-\ref{rmk:rank2UN_2}.\par
The superconformal index $\mI_{\mathfrak{su}}$ of such a quiver gauge theory takes the form \eqref{eq:ZPExp}, with the integration domain $G^{\prime}$ replaced by the gauge group $G$ \eqref{eq:Ggauge}, and integrand 
\begin{equation}\label{eq:fXSCI}
\begin{aligned}
    f(X) = \PE & \left[ \sum_{i\in\ver} \left[ \sli_{V}(\mathbf{p})\left(\sum_{a,b=1}^{N_i}x_{a,i}x_{b,i}^{-1}-1\right) +\sli_{H}(\mathbf{p})\chi_{F_i} (\mathbf{y})\sum_{a=1}^{N_i}\left(x_{a,i}+x_{a,i}^{-1}\right)\right] \right. \\
    & \left. + \sum_{\substack{(i,j)\in\ver^2 \\ i \le j }}\sum_{\{ e: i \to j \}\subset \edge } \sli_{H} (\mathbf{p})  \sum_{a=1}^{N_i}\sum_{b=1}^{N_j}\left(x_{a,i}x_{b,j}^{-1}+x_{a,i}^{-1}x_{b,j}\right)  \right] .
\end{aligned}
\end{equation}
The plethystic exponential is defined as 
\begin{equation}\label{eq:plethystic exp}
    \PE\left[f(t_1,\cdots ,t_r)\right]=\exp\left(\sum_{n\geq 1}\frac{1}{n}f\left(t_1^n,\cdots ,t_r^n\right)\right) .
\end{equation}
The functions $\sli_{V} (\mathbf{p}), \sli_{H} (\mathbf{p})$ in \eqref{eq:fXSCI} are the single-letter indices, and $\mathbf{p}$ collectively denotes the fugacities for spacetime and R-symmetries. Besides, $\chi_{F_i} (\mathbf{y})$ is the character of the fundamental representation of the $i^{\text{th}}$ flavour group, evaluated on the flavour fugacities $\mathbf{y}$.\footnote{To lighten the expressions, we turn off flavour fugacities except for the fundamental hypermultiplets. They can be straightforwardly reinserted in the coefficients $h^{e}_n$.}
Comparing with \eqref{eq:fMM}, we have
\begin{equation}\label{eq:IfromZ}
    \mI_{\mathfrak{su}}= \exp \left( -\sum_{i\in\ver} \sum_{n \ge 1} \frac{h^{ii}_n}{n}\right) \left.\mI\right\rvert_{G^{\prime} \mapsto G} ,
\end{equation}
provided the identifications 
\begin{align}
    v^{i}_n &= \sli_{H}(\mathbf{p}^n)\chi_{F_i} (\mathbf{y}^n) \label{eq:Indexvi}\\
    h^{V,i}_n &= \sli_{V}(\mathbf{p}^n) \\
    h^{e}_n &=\sli_{H} (\mathbf{p}^n) 
\end{align}
from which the matrices \eqref{eq:hijgeneric} are given by 
\begin{align}
    h^{ij}_n &= \delta^{ij}\sli_{V}(\mathbf{p}^n) +\sum_{\{ e: i \to j \}} \sli_{H} (\mathbf{p}^n) ,\qquad n \in \Z_{\geq 1}. \label{eq:Indexhij}
\end{align}

\begin{rmk}[(Anti-)Symmetric hypermultiplets]\label{rmk:rank2UN_1}
Adding $S_i$ hypermultiplets in the rank-2 symmetric and $\Lambda_i$ hypermultiplets in the antisymmetric representation of $SU(N_i)$, the argument of the plethystic exponential in \eqref{eq:fXSCI} acquires additional terms 
\begin{equation}
    \sum_{i\in\ver} \left[ S_i \sli_{H} (\mathbf{p})  \sum_{1 \le a \le b \le N_i} \left(x_{a,i}x_{b,i}+x_{a,i}^{-1}x_{b,i}^{-1}\right) + \Lambda_i \sli_{H} (\mathbf{p})  \sum_{1 \le a < b \le N_i} \left(x_{a,i}x_{b,i}+x_{a,i}^{-1}x_{b,i}^{-1}\right) \right].
\end{equation}
After elementary manipulations, \eqref{eq:SeffGeneral} is modified by the addition of the terms
\begin{equation}
    \sum_{n\geq 1}\frac{\sli_{H} (\mathbf{p}^n) }{n} \sum_{i\in\ver} \left[ \left( S_i + \Lambda_i \right)\sum_{a=1}^{N_i}\sum_{b=1}^{N_i} \cos (n(\theta_{a,i}+\theta_{b,i})) +\left( S_i - \Lambda_i \right) \sum_{a=1}^{N_i} \cos (2n \theta_{a,i}) \right].
\end{equation}
\end{rmk}

\subsection{Large \texorpdfstring{$N$}{N} limit for general quiver matrix model}\label{sec:LargeNMM}
We now study the large $N$ regime of the matrix models in section \ref{sec:QuiverMM} (see \cite{Santilli:2025zum} for a review of large $N$ limits). In order to take the limit of large $N_j$ uniformly, we introduce a parameter $N\in \mathbb{N}$ and eventually take $N\to \infty$ in such a way that $\frac{N_j}{N}= \mathcal{O}(1)$.\par
Introduce the eigenvalue densities
\begin{equation}\label{eq:defrhoi}
    \rho_i(\theta)=\frac{1}{N}\sum_{a=1}^{N_i} \delta(\theta-\theta_{a,i})\;, \qquad \forall i \in \ver ,
\end{equation}
which are subject to the normalisation condition
\begin{equation}\label{eq:normalisationrho}
    \int_{-\pi}^{\pi}\dd\theta\rho_i(\theta)=\frac{N_i}{N}\;.
\end{equation}
Defining the matrix 
\begin{equation}\label{eq:hmatrix}
    (h_n-\mathbb{I})^{ij} := h_n^{ij} - \delta^{ij} ,
\end{equation}
the effective action \eqref{eq:SeffGeneral} is expressed as
\begin{equation}\label{eq:Seffrho2}
\begin{aligned}
    S_{\mathrm{eff}}=N^2\sum_{n\geq 1}\frac{1}{n} & \sum_{i\in\ver}\int_{-\pi}^{\pi}\dd\theta_1\rho_i(\theta_1) \left[ \frac{2v^{i}_n}{N} \cos(n\theta_{1})\right. \\ 
    +&\left. \sum_{j\in\ver}(h_n-\mathbb{I})^{ij} \int_{-\pi}^{\pi} \dd \theta_2 \rho_j(\theta_2)\cos(n(\theta_{1}-\theta_2))\right] .
\end{aligned}
\end{equation}\par
Next, we introduce the Fourier coefficients of the eigenvalue densities:
\begin{equation}
    \chi_{j,n} = \frac{1}{\pi}\int_{-\pi}^{\pi} \dd\theta\rho_j(\theta) \cos(n\theta) , \qquad \psi_{j,n} = \frac{1}{\pi}\int_{-\pi}^{\pi} \dd\theta\rho_j(\theta) \sin(n\theta) .
\end{equation}
The $0^{\text{th}}$ Fourier mode is fixed by normalisation \eqref{eq:normalisationrho}.
Then, \eqref{eq:Seffrho2} becomes:
\begin{equation}\label{eq:Seffrho3}
\begin{aligned}
    S_{\mathrm{eff}}&=\pi^2 N^2\sum_{n\geq 1}\frac{1}{n} \sum_{i\in\ver} \left[ \frac{2v^{i}_n}{\pi N} \chi_{i,n} + \sum_{j\in\ver}(h_n-\mathbb{I})^{ij} \left( \chi_{i,n} \chi_{j,n} + \psi_{i,n} \psi_{j,n}\right)\right] .
\end{aligned}
\end{equation}
The saddle point equations obtained from this general action read
\begin{align}
    \frac{v^{i}_n}{\pi N} + \sum_{j \in \ver} (h_n-\mathbb{I})^{ij} \chi_{j,n} &= 0 , \label{eq:SPE1}\\
    \sum_{j \in \ver} (h_n-\mathbb{I})^{ij} \psi_{j,n} &= 0 ,\label{eq:SPE2}
\end{align}
for all $i \in \ver$ and for every $n \ge 1$ (see for instance \cite{Santilli:2025zum} for more details on the derivation).\par 
Observe that, on the solutions to the system \eqref{eq:SPE1}-\eqref{eq:SPE2}, we find
\begin{equation}\label{eq:SeffSPE}
    \left.S_{\mathrm{eff}} \right\rvert_{\text{saddle}}= N \pi \sum_{i\in\ver} \sum_{n\geq 1}\frac{1}{n} v^{i}_n \chi_{i,n} .
\end{equation}
In the next sections we will solve \eqref{eq:SPE1}-\eqref{eq:SPE2} and compute \eqref{eq:SeffSPE} first for finite quivers of arbitrary shape in section \ref{sec:FiniteL}, and then for quivers with $\lvert \ver\rvert \to \infty$ in section \ref{sec:LargeL}.

\begin{rmk}[(Anti-)Symmetric hypermultiplets]\label{rmk:rank2UN_2}
The modifications to include symmetric or antisymmetric hypermultiplets have been spelled out in Remark \ref{rmk:rank2UN_1}. These are incorporated in \eqref{eq:SPE1}-\eqref{eq:SPE2} as follows: in \eqref{eq:SPE1}, $h_n^{ii} \mapsto  \sli_{V}(\mathbf{p}^n) + (g_i+S_i+\Lambda_i) \sli_{H} (\mathbf{p}^n)$, and, when $n$ is even, $v^{i}_n \mapsto \sli_{H}(\mathbf{p}^n)\chi_{F_i} (\mathbf{y}^n) + (S_i-\Lambda_i) \sli_{H} (\mathbf{p}^{n/2})$; in \eqref{eq:SPE2}, $h_n^{ii} \mapsto  \sli_{V}(\mathbf{p}^n) + (g_i-S_i-\Lambda_i) \sli_{H} (\mathbf{p}^n) $.
\end{rmk}

\section{Universal solution for large \texorpdfstring{$N$}{N} quivers}
\label{sec:FiniteL}
In this section we solve the saddle point equations \eqref{eq:SPE1}-\eqref{eq:SPE2} for quivers of arbitrary shape.\par
The solution to \eqref{eq:SPE2} is $\psi_{j,n}=0$ $\forall n \in \N$ and $\forall j \in \ver$. Physically, this follows from the fact that we only consider theories with `vector' (non-chiral) matter content. This in particular implies that the saddle point eigenvalue configurations are traceless, $\int_{-\pi}^{\pi} \theta \rho_i (\theta) \dd \theta =0 \forall i \in \ver$.\par
Next, for every $n \in \N$ we introduce a matrix $D_n$, defined as the inverse matrix of the negative of \eqref{eq:hmatrix}, that is,
\begin{equation}
    \left[ D_n \right]_{ij} = \left[\left( \mathbb{I}- h_n\right)^{-1}\right]_{ij} , \qquad \forall i,j \in \ver .
\end{equation}
We are assuming that $\det (\mathbb{I}- h_n)\ne 0$, which is guaranteed to hold for generic fugacities. We elaborate on the physical meaning of this hypothesis in section \ref{sec:Hagedorn}. Then, \eqref{eq:SPE1} is solved by 
\begin{equation}
    \chi_{i,n} = \frac{1}{\pi N} \sum_{j \in \ver}  \left[ D_n \right]_{ij} v^{j}_n .
\end{equation}\par
Lastly, one can show that $\lim_{N\to\infty} \frac{1}{C}\exp\left(\sum_{j\in\ver}N_j \sum_{n\geq 1} \frac{1}{n}\right) =1$, thus the overall divergent factor in \eqref{eq:Zeff} (originated from the Haar measure) cancels against the normalisation.\par

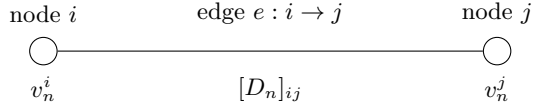
\begin{figure}[th]
    \centering
    \begin{tikzpicture}
        \node[circle,draw] (i) at (-3,0) { \hspace{10pt}};
        \node[circle,draw] (j) at (3,0) { \hspace{10pt}};
        \draw[-] (i) -- (j);
        \node at (-3,0.5) {\footnotesize node $i$};
        \node at (3,0.5) {\footnotesize node $j$};
        \node at (0,0.5) {\footnotesize edge $e:i \to j$};
        \node at (-3,-0.5) {\footnotesize $v^i_{n}$};
        \node at (3,-0.5) {\footnotesize $v^j_{n}$};
        \node at (0,-0.5) {\footnotesize $[D_{n}]_{ij}$};
    \end{tikzpicture}
    \caption{Schematic representation of the $n^{\text{th}}$ contribution to \eqref{eq:Zuniversal}.}
    \label{fig:propagatorij}
\end{figure}\par

Evaluating the quiver matrix model at large $N$ we find the universal formula 
\begin{equation}\label{eq:Zuniversal}
    \left.\ln \mI \right\rvert_{N\to\infty} = \sum_{n=1}^{\infty}\left(  \frac{1}{n} \sum_{i,j \in \ver} \left[ D_n \right]_{ij} v^{i}_n v^{j}_n  +\ln \det_{i,j \in \ver}\left[ D_n \right]_{ij} \right).
\end{equation}
The first piece is the on-shell effective action \eqref{eq:SeffSPE}, and the logarithm of the determinant originates from the usual Gaussian integration over fluctuations of all $\chi_{j,n}, \psi_{j,n}$ around their saddle point values.\footnote{The integration measure over the infinitely many Fourier modes is normalised compatibly with the normalisation of \eqref{eq:ZPExp}, so that $\mI=1+\mathcal{O}(\mathbf{p})$.} The structure of \eqref{eq:Zuniversal} is schematically depicted in figure \ref{fig:propagatorij}.

\subsection{Supersymmetric index of arbitrary special unitary quivers}
\label{sec:IndexSUresult}
By \eqref{eq:IfromZ}, it suffices to plug \eqref{eq:Indexvi}-\eqref{eq:Indexhij} into the universal formula \eqref{eq:Zuniversal} to obtain the supersymmetric index of any 4d theory with 8 supercharges, or the  perturbative index of any 5d theory with 8 supercharges. We find:
\begin{equation}\label{eq:Iuniversal}
    \left.\mI_{\mathfrak{su}} \right\rvert_{N\to\infty} = \frac{ \exp \left\{\sum_{n=1}^{\infty} \frac{1}{n} \left[ \sli_{H}(\mathbf{p}^n)^2 \sum_{i,j \in \ver} \left[\left( \mathbb{I}- h_n\right)^{-1}\right]_{ij} \chi_{F_i}(\mathbf{y}^n)\chi_{F_j}(\mathbf{y}^n) -\tr \left( h_n\right)\right]  \right\}}{\prod_{n=1}^{\infty}\det \left( \mathbb{I}- h_n \right) },
\end{equation}
where 
\begin{equation}
    h^{ii}_n = \sli_{V}(\mathbf{p}^n) + g_i \sli_{H} (\mathbf{p}^n) , \qquad  \left.h^{ij}_n\right\rvert_{i \ne j} = \sum_{\{ e: i \to j \} \subset \edge} \sli_{H} (\mathbf{p}^n) .
\end{equation}
Based on Remark \ref{rmk:rank2UN_2}, we can immediately generalise the result to include an arbitrary number of hypermultiplets in the symmetric or antisymmetric representation, obtaining:
\begin{equation}\label{eq:Iuniv2}
\begin{aligned}
    \left.\mI_{\mathfrak{su}} \right\rvert_{N\to\infty} &= \exp \left\{ \sli_{H}(\mathbf{p}^n)^2 \sum_{i,j \in \ver} \left[ \sum_{\substack{ n \ge 1\\ n \text{ odd}}} \frac{1}{n} \left[\left( \mathbb{I}- h_n^{(+)}\right)^{-1}\right]_{ij} \chi_{F_i}(\mathbf{y}^n)\chi_{F_j}(\mathbf{y}^n) \right.\right. \\
    & \left. + \sum_{\substack{ n \ge 1\\ n \text{ even}}} \frac{1}{n} \left[\left( \mathbb{I}- h_n^{(+)}\right)^{-1}\right]_{ij}  \left( \chi_{F_i}(\mathbf{y}^n) +(S_i- \Lambda_i) \frac{\sli_{H} (\mathbf{p}^{\frac{n}{2}})}{ \sli_{H}(\mathbf{p}^n)} \right)\left( \chi_{F_j}(\mathbf{y}^n) + (S_j- \Lambda_j)\frac{\sli_{H} (\mathbf{p}^{\frac{n}{2}}) }{ \sli_{H}(\mathbf{p}^n)} \right) \right] \\
    & \left. + \sum_{n\ge 1} \left[ -\frac{1}{n} \tr \left( h_n\right) + \ln \sqrt{\det\left(\mathbb{I}- h_n^{(+)}\right)\det\left(\mathbb{I}- h_n^{(-)}\right) } \right]  \right\},
\end{aligned}
\end{equation}
with $ h_n^{(\pm), ii} = h^{ii}_n \pm \left( S_i \sli_{H} (\mathbf{p}^{n}) + \Lambda_i \sli_{H} (\mathbf{p}^{n})\right)$, and the rest as above.\par
\begin{rmk}[4d $\mathcal{N}=1$ gauge theories]
\label{rmk:4dN1index}
    The derivation does not rely on having 8 supercharges, but only on having vector matter content: every edge of the graph, representing a hypermultiplet, is replaced by a pair of opposite arrows, representing a pair of chiral/anti-chiral multiplets. Repeating our procedure, one arrives at the same universal solution \eqref{eq:Iuniversal} for the supersymmetric index of any 4d $\mN=1$ quiver gauge theory. The only minor differences are in the single-letter indices.
    The final result matches the independent derivation of Fang--Feng--Xie \cite{friends}.
\end{rmk}
As two quick consistency checks, we observe:
\begin{itemize}
    \item For theories with a single gauge group $SU(N)$, \eqref{eq:Iuniversal} reproduces the classic result of \cite{Dolan:2008qi};
    \item For theories without fundamental hypermultiplets, \eqref{eq:Iuniversal} simplifies into 
    \begin{equation}\label{eq:ISUnofund}
        \left.\mI_{\mathfrak{su}} \right\rvert_{N\to\infty} = \frac{\exp \left( - \sum_{n=1}^{\infty} \frac{1}{n}\tr (h_n) \right) }{\prod_{n=1}^{\infty} \det \left( \mathbb{I} - h_n \right)} ,
    \end{equation}
    which reproduces the result of \cite{Gadde:2010en}, and extends it to quivers with adjoint hypermultiplets.
\end{itemize}\par

\subsection{Supersymmetric index of arbitrary orthosymplectic quivers}
\label{sec:IndexSOSpresult}
We now extend the derivation of the index of arbitrary quivers to gauge theories with gauge group 
\begin{equation}\label{eq:SOSpgauge}
    G=\prod_{i\in\ver} G_i, \qquad G_i \in \left\{ SO(2N_i), SO(2N_i+1), USp(2N_i)\right\} \;,
\end{equation}
and $F_i$ half-hypermultiplets in the fundamental representation, $S_i$ rank-2 symmetric, $\Lambda_i$ rank-2 antisymmetric half-hypermultiplets, and an arbitrary number of bifundamental half-hypermultiplets represented by edges $e:i \to j$.\par
The derivation is very similar to the one in section \ref{sec:General}, albeit more cumbersome; we defer it to appendix \ref{app:SOSpderivation}, and state the result here. 
We find the supersymmetric index $\mI_{\mathfrak{osp}}$ of a quiver of arbitrary shape and gauge group \eqref{eq:SOSpgauge} to be
\begin{equation}\label{eq:ISOSpuniversal}
    \left.\mI_{\mathfrak{osp}} \right\rvert_{N\to\infty} = \frac{ \exp \left\{\sum_{n=1}^{\infty} \frac{1}{n} \left[ \frac{1}{2}\sum_{i,j \in \ver} \left[\left( \mathbb{I}- h_n \right)^{-1}\right]_{ij} \tilde{v}^{i}_n \tilde{v}^{j}_n +c_n \right] \right\}}{\prod_{n=1}^{\infty}\sqrt{\det \left( \mathbb{I}- h_n\right) }},
\end{equation}
provided \eqref{eq:wfromvh} and the identifications \eqref{eq:SOSpIndexvi}-\eqref{eq:SOSpIIndexhij} below, which we now explain.\par
In \eqref{eq:ISOSpuniversal}, we have used the symmetry $h_n^{ij}=h_n^{ji}$, and $c_n$ collects factored-out contributions from the trivial eigenvalues in the maximal tori of all $SO(2N_i+1)$.
To write the subsequent expressions in a more compact form, let 
\begin{equation}
\delta_{n \in 2 \Z}= \begin{cases} 1 & n \text{ even} \\ 0 & n \text{ odd}, \end{cases}   
\end{equation} 
and define the coefficients
\begin{equation}\label{defshiftSOSp}
	\kappa_j = \begin{cases} 1 & G_j=SO(2N_j+1) \\ 0 & \text{otherwise}, \end{cases} \qquad \sigma^{j}_{n} = \begin{cases}- \delta_{n \in 2 \Z} & G_j = SO(2N_j) \\ 1-\delta_{n \in 2 \Z} & G_j = SO(2N_j+1) \\ \delta_{n \in 2 \Z} & G_j = USp(2N_j).\end{cases} 
\end{equation}
We can then express $\tilde{v}^{i}_n$ in \eqref{eq:ISOSpuniversal} as
\begin{equation}\label{eq:wfromvh}
    \tilde{v}^{i}_n = v^{i}_n - \sigma^{i}_{n} + \sum_{j\in\ver} \kappa_j h^{ij}_n  .
\end{equation}
We then write $v^{i}_n = v^{i}_{\Box , n} + v^{i}_{\circlearrowleft, n}$, splitting the contributions to $v^{i}_n$ from the framing and loops of the quiver. To turn the universal large $N$ formula \eqref{eq:ISOSpuniversal} into an index we plug
\begin{align}
    v^{i}_{\Box,n} &= \sli_{H}(\mathbf{p}^n) \chi_{F_i} (\mathbf{y}^n)  \label{eq:SOSpIndexvi}\\
    v^{i}_{\circlearrowleft, n} &= \begin{cases} \delta_{n \in 2 \Z} \left[ (S_i -\Lambda_i)\sli_{H}(\mathbf{p}^{n/2}) + \sli_{V}(\mathbf{p}^{n/2}) \right] & G_i = USp(2N_i)  \\ 
    \delta_{n \in 2 \Z} \left[ (S_i -\Lambda_i)\sli_{H}(\mathbf{p}^{n/2}) - \sli_{V}(\mathbf{p}^{n/2}) \right] & G_i = SO(2N_i) \\
    \delta_{n \in 2 \Z} \left[ (S_i -\Lambda_i)\sli_{H}(\mathbf{p}^{n/2}) - \sli_{V}(\mathbf{p}^{n/2}) \right]  & G_i = SO(2N_i+1)  \end{cases} \\
    h^{ii}_n &= \sli_{V}(\mathbf{p}^n) + (S_i + \Lambda_i) \sli_{H} (\mathbf{p}^n) \label{eq:SOSpIIndexhii}\\
    h^{ij}_n &=\sum_{\{ e: i \to j \}} \sli_{H} (\mathbf{p}^n) ,\qquad j\ne i . \label{eq:SOSpIIndexhij}
\end{align}
Finally we have
\begin{equation}\label{eq:cnshiftSOodd}
    c_n = \sum_{i \in \ver} \kappa_i \sli_{H}(\mathbf{p}^n) \left( \chi_{F_i} (\mathbf{y}^n) + S_i\right) + \frac{1}{2}\sum_{\substack{ (i,j)\in \ver^2 \\ i \ne j }} \kappa_i \kappa_j h^{ij}_n .
\end{equation}

\begin{rmk}\label{rmk:irrepPESOSp}
    The rank-2 antisymmetric representation of $USp(2N)$ is reducible, and decomposes into the direct sum of the traceless rank-2 antisymmetric and the trivial representation. To get the index with the free half-hypermultiplets discarded, one should divide by $\PE \left[ \Lambda_i \sli_H(\mathbf{p})\right]$ for $G_i=USp(2N_i)$. Likewise, the rank-2 symmetric representation of the special orthogonal group is reducible, and one should divide by $\PE \left[ S_i \sli_H(\mathbf{p})\right]$ for $G_i=SO(2N_i), SO(2N_i+1)$.
\end{rmk}
\begin{rmk}[4d $\mathcal{N}=1$ gauge theories]
    As for the unitary case, the derivation extends directly to 4d $\mN=1$ theories with vector matter content. We initially obtained the result for 4d $\mN=1$ theories with $G_i \in \left\{ SO(2N_i),USp(2N_i)\right\}$, and extended it to full generality motivated by the work of Fang--Feng--Xie \cite{friends}. The final result matches their independent derivation.
\end{rmk}

\section{Long quiver limit}
\label{sec:LargeL}
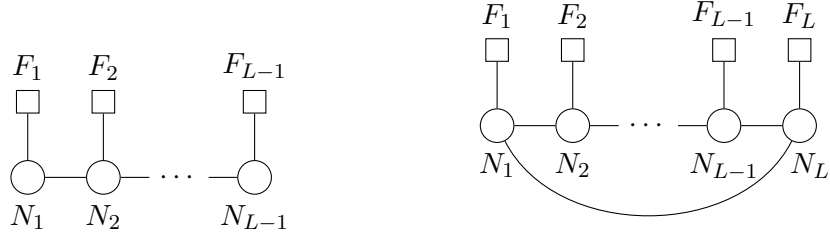
\begin{figure}[htb]
\centering
         
             \begin{tikzpicture}
                      \node[label=below:{$N_1$}][u](N1){ \hspace{8pt} };
                      \node[label=above:{$F_1$}][uf](F1)[above of=N1]{ \hspace{12pt}};
                      \node[label=below:{$N_2$}][u](N2)[right of=N1]{ \hspace{8pt} };
                      \node[label=above:{$F_2$}][uf](F2)[above of=N2]{ \hspace{12pt}};
                      \node (dots)[right of=N2]{$\dotsb$};
                      \node[label=below:{$N_{L-1}$}][u](NP-1)[right of=dots]{ \hspace{8pt} };
                      \node[label=above:{$F_{L-1}$}][uf](FP-1)[above of=NP-1]{ \hspace{12pt}};
                      \draw(N1)--(F1);
                      \draw(N1)--(N2);
                      \draw(N2)--(F2);
                      \draw(N2)--(dots);
                      \draw(dots)--(NP-1);
                      \draw(NP-1)--(FP-1);
             \end{tikzpicture}
\hspace{2cm}
             \begin{tikzpicture}
                      \node[label=below:{$N_1$}][u](N1){ \hspace{8pt} };
                      \node[label=above:{$F_1$}][uf](F1)[above of=N1]{ \hspace{12pt}};
                      \node[label=below:{$N_2$}][u](N2)[right of=N1]{ \hspace{8pt} };
                      \node[label=above:{$F_2$}][uf](F2)[above of=N2]{ \hspace{12pt}};
                      \node (dots)[right of=N2]{$\dotsb$};
                      \node[label=below:{$N_{L-1}$}][u](NP-1)[right of=dots]{ \hspace{8pt} };
                      \node[label=above:{$F_{L-1}$}][uf](FP-1)[above of=NP-1]{ \hspace{12pt}};
                      \node[label=below:{\hspace{8pt}$N_{L}$}][u](NP)[right of=NP-1]{ \hspace{8pt} };
                      \node[label=above:{$F_{L}$}][uf](FP)[above of=NP]{ \hspace{12pt}};
                      \draw(N1)--(F1);
                      \draw(N1)--(N2);
                      \draw(N2)--(F2);
                      \draw(N2)--(dots);
                      \draw(dots)--(NP-1);
                      \draw(NP-1)--(FP-1);
                      \draw(NP-1)--(NP);
                      \draw(FP)--(NP);
                       \draw (N1)to[out=-63, in=-117](NP);
             \end{tikzpicture}

\caption{Linear, and circular quivers describing a supersymmetric gauge theory with 8 supercharges. The $i^{\text{th}}$ circular node represents a vector multiplet in the adjoint representation of $SU(N_i)$. The square node attached to the $i^{\text{th}}$ circular node represents a flavour symmetry $\mathfrak{u} (F_i)$. The edges connecting neighbouring gauge nodes represent hypermultiplets transforming in the bifundamental representation of $SU(N_i)\times SU(N_{i+1})$.}
\label{fig:quiver}
\end{figure}

In this section, we specialise to linear, or circular quivers. The corresponding graph is the same as the Dynkin diagram of the $A_{L-1}$, or the affine $\hat{A}_{L-1}$ algebra, as depicted in figure \ref{fig:quiver}.\par
These theories have a parameter $L\sim\lvert \ver \rvert$, which can be taken to be large. The goal of this section is to evaluate the limit of large $N$, and large $L$, which shall be referred to as the \emph{long quiver limit}. 

For these theories, $v^{i}_n=F_i i_H(\mathbf{p}^n)$ from \eqref{eq:Indexvi}, and from \eqref{eq:Indexhij} 
\begin{equation}\label{eq:hijlinear}
    h_n^{ij} = \begin{cases} \sli_V (\mathbf{p}^n) & j=i \\ \sli_H (\mathbf{p}^n) & j=i\pm 1 \\ 0 & \text{otherwise},\end{cases}
\end{equation}
where the indices are understood $\mod L$ in the affine $\hat{A}_{L-1}$ case. The saddle point equations \eqref{eq:SPE1}-\eqref{eq:SPE2} read
\begin{align}
    \frac{F_i}{\pi N} + \left( \frac{\sli_V (\mathbf{p}^n) -1}{\sli_H (\mathbf{p}^n)} \right)\chi_{i,n} + \chi_{i+1,n}+ \chi_{i-1,n} &= 0 , \\
    \left( \frac{\sli_V (\mathbf{p}^n) -1}{\sli_H (\mathbf{p}^n)} \right)\psi_{i,n} + \psi_{i+1,n}+ \psi_{i-1,n} &= 0 .
\end{align}
One must equip this system of equations with the appropriate boundary conditions
\begin{equation}\label{eq:LongQbc}
     \begin{cases} \chi_{0,n}=0=\chi_{L,n} , \quad \psi_{0,n}=0=\psi_{L,n} , & A_{L-1} ,\\ \chi_{0,n}=\chi_{L,n} , \quad\qquad \psi_{0,n}=\psi_{L,n} , & \hat{A}_{L-1} ,\end{cases}
\end{equation}
for linear and circular quivers respectively. While the saddle point equations at the nodes $i\in\{0,L\}$ for the linear case would, a priori, require a separate treatment, it is easy to check that they can be presented in the current form provided the boundary conditions \eqref{eq:LongQbc}.\par
The large $L$ limit is then obtained by introducing the effective coordinate $z\in[0,1]$, which indicates the position of a node in question along the quiver. One then promotes the eigenvalue densities to a continuous function of $z$, as
\begin{equation}
    \chi_{i,n}\mapsto \chi_n(z)\;, \qquad \psi_{i,n}\mapsto \psi_n(z)\;.
\end{equation}
We also define the flavour rank function, 
normalised as\footnote{Reinserting flavour fugacities, $\varphi(z)$ acquires a very simple dependence on $n$ through $F_i \mapsto \chi_{F_i}(\mathbf{y}^n)$.}
\begin{equation}\label{eq:flavour rank function}
    \varphi (z) = \frac{1}{NL} \sum_{i=1}^{L} F_i \delta \left( z- \frac{i}{L} \right) .
\end{equation}
The neighbouring-node interaction is expanded using the continuum relation \cite{Akhond:2022oaf}
\begin{equation}\label{eq:continuum limit difference}
    \chi_{i+1,n} + \chi_{i-1,n} = 2 \chi_{i,n}  + \left(\chi_{i+1,n} - \chi_{i,n}\right)- \left(\chi_{i,n} - \chi_{i-1,n}\right) \mapsto 2 \chi_{i,n} + \frac{1}{L^2} \ddot{\chi}_n (z) ,
\end{equation}
and likewise for $\psi_n (z)$, which follows from the Taylor expansion of the density profile along the quiver direction. Here and in the following, $\dot{f} = \frac{\partial f}{\partial z}$. The effective action in the long quiver limit then reads
\begin{equation}\label{eq:SeffSPELongQ}
\begin{aligned}
    S_{\mathrm{eff}}=LN^2\pi^2\int_0^1\dd z\sum_{n\geq 1}\frac{1}{n}\bigg[&\left(\sli_V(\mathbf{p}^n)+2\sli_H(\mathbf{p}^n)-1\right)\left(\psi_n^2+\chi_n^2\right)+\frac{2}{\pi N}\sli_H(\mathbf{p}^n)\varphi(z)\chi_n\\&+\frac{1}{L^2}\sli_H(\mathbf{p}^n)\left(\psi_n\ddot{\psi}_n+\chi_n\ddot{\chi}_n\right)\bigg] \, ,
    \end{aligned}
\end{equation}
after dropping the fugacity-independent term that cancels against the normalisation. This gives the long quiver limit of the saddle point equations 
\begin{align}
     \left( \frac{\sli_V (\mathbf{p}^n)+2\sli_H (\mathbf{p}^n) -1}{\sli_H (\mathbf{p}^n)} \right)\chi_{n}(z) + \frac{1}{L^2} \ddot{\chi}_n (z) &= -\frac{1}{\pi}\varphi (z) , \label{eq:LongQSPE1}\\
    \left( \frac{\sli_V (\mathbf{p}^n)+2\sli_H (\mathbf{p}^n) -1}{\sli_H (\mathbf{p}^n)} \right)\psi_{n}(z) + \frac{1}{L^2} \ddot{\psi}_n (z)  &= 0 .\label{eq:LongQSPE2}
\end{align}
We can immediately plug these conditions back into \eqref{eq:SeffSPELongQ} to obtain 
\begin{equation}
    \left.S_{\mathrm{eff}} \right\rvert_{\text{saddle}}=LN^2 \pi \sum_{n\geq 1}\frac{\sli_H (\mathbf{p}^n)}{n} \int_0^{1}\dd z \varphi(z) \chi_{n} (z) .
\end{equation}
The remaining task is to find the solution to \eqref{eq:LongQSPE1}, subject to the appropriate boundary conditions \eqref{eq:LongQbc}. To this end, consider the one-dimensional Green's function, satisfying
\begin{equation}
  -  \left(\frac{1}{L^2}\partial_z{}^2+\frac{\sli_V(\mathbf{p}^n)+2\sli_H(\mathbf{p}^n)-1}{\sli_H(\mathbf{p}^n)}\right)G_n(z,z')=\delta(z-z')\;.
\end{equation}
The solution to the saddle-point equation is then (see e.g. \cite[Ch.V.14]{CourantHilbertBook})
\begin{equation}
    \left.\chi_n (z)\right\rvert_{\mathrm{saddle}}=\frac{1}{\pi}\int_0^1 G_n(z,z')\varphi(z') \dd z'\;,
\end{equation}
under the mild assumption that $\varphi (z)$ is well-approximated by a piecewise continuous function in the limit. 
The on-shell action can therefore be expressed as
\begin{equation}\label{eq:on-shell action long quiver}
    \left.S_{\mathrm{eff}}\right\rvert_{\mathrm{saddle}}=LN^2 \sum_{n\geq 1}\frac{\sli_H (\mathbf{p}^n)}{n}\int_0^1 \int_0^1 \varphi(z)G_n(z,z')\varphi(z')\dd z\dd z' .
\end{equation}\par
\medskip
It is convenient to define $t_n$ via the relation 
\begin{equation}\label{eq:deftnHL}
    t_n+\frac{1}{t_n} = \frac{ 1-\sli_V (\mathbf{p}^n)}{\sli_H (\mathbf{p}^n)} .
\end{equation}
We also introduce the effective masses
\begin{equation}\label{eq:Greenmass}
    M_n = \sqrt{-\frac{i_V(\mathbf{p}^n)+2i_H(\mathbf{p}^n)-1}{i_H(\mathbf{p}^n)} } = \left( t_n^{-\frac{1}{2}} - t_n^{\frac{1}{2}} \right)  .
\end{equation}
The Green's functions $G_n$ in the linear and circular quivers are explicitly given.
\begin{itemize}
    \item For linear $A_{L-1}$ quivers, imposing Dirichlet boundary conditions we have 
    \begin{equation}\label{eq:Gnlinear}
    \begin{aligned}
        G_n (z,z') = \frac{L}{M_n \sinh (M_n L)}\sinh(M_n L\,z_<)\sinh\left(M_n L(1-z_>)\right)\;,
    \end{aligned}
    \end{equation}
    where $z_<=\min(z,z')$, and $z_>=\max(z,z')$
    \item For circular $\hat{A}_{L-1}$ quivers, we impose periodic boundary conditions and get 
    \begin{equation}
        G_n(z,z') = \,\frac{L}{2M_n}\,\frac{\cosh\!\Big(M_n L\big(\tfrac{1}{2}-|z-z'|\big)\Big)}{\sinh\!\left(\dfrac{M_n L}{2}\right)}.
    \end{equation}
\end{itemize}

\subsection{Exact evaluation of the long quiver index}
\label{sec:exactILong}
Integrating out the Gaussian fluctuations around the saddle point, we arrive at 
\begin{equation}
    \left.\mI_{\mathfrak{su}} \right\rvert_{\substack{N \to \infty \\ L \to \infty}} = \frac{ \exp \left(  \left.S_{\mathrm{eff}}\right\rvert_{\mathrm{saddle}} \right)}{\prod_{n\geq1}\det \left( \mathbb{I}- h_n \right) } .
\end{equation}
The on-shell action is given in \eqref{eq:on-shell action long quiver}, and for $h_n$ in \eqref{eq:hijlinear} the determinant is computed as 
\begin{equation}
    \det \left( \mathbb{I}- h_n \right) = \prod_{j=1}^{L-1}\left( 1-\sli_V (\mathbf{p}^n) -2 \sli_H (\mathbf{p}^n) \cos \frac{j \pi}{L}\right) .
\end{equation}
In the long quiver limit, we write 
\begin{equation}
    \prod_{n\geq 1}\frac{1}{\det \left( \mathbb{I}- h_n \right) } = \exp \left\{ -\sum_{n\geq1} L \left[ \ln \sli_H (\mathbf{p}^n) + \int_0^1\dd z\ln \left( \frac{ 1-\sli_V (\mathbf{p}^n)}{\sli_H (\mathbf{p}^n)} -2 \cos (\pi z) \right) \right]\right\} .
\end{equation}
The integral can be evaluated exactly. Using the definition \eqref{eq:deftnHL} and plugging the flavour rank function \eqref{eq:flavour rank function} into \eqref{eq:on-shell action long quiver}, we finally arrive at 
\begin{equation}\label{eq:IlongQ}
    \left.\mI_{\mathfrak{su}} \right\rvert_{\substack{N \to \infty \\ L \to \infty}} = \exp \left\{ L\sum_{n\geq 1} \left[\frac{\sli_H (\mathbf{p}^n)}{n} L^{-2}\sum_{i,j=1}^{L} F_iF_j G_n\left( \frac{i}{L}, \frac{j}{L}\right)- \frac{\sli_V (\mathbf{p}^n)}{n}  + \ln \left(\frac{t_n}{\sli_H (\mathbf{p}^n)}\right)\right] \right\} .
\end{equation}
In the scaling limit with $F_i = \mathcal{O}(N)$, the second and third terms are subleading, and, taking the plethystic logarithm of \eqref{eq:IlongQ} we may obtain an exact expression
\begin{equation}\label{eq:PLLongQ}
    \left.\mathrm{PL}[\mI_{\mathfrak{su}} ]\right\rvert_{\substack{N \to \infty \\ L \to \infty}}=L^{-1} \sli_H(\mathbf{p}) \sum_{i,j=1}^L F_i F_jG_1\left( \frac{i}{L}, \frac{j}{L}\right)\;.
\end{equation}

\subsection{Hall--Littlewood limit and a first consistency check}
It is insightful to examine the Hall--Littlewood limit \cite{Gadde:2011uv} of our solution in 4d $\mN=2$ theories. For theories with complete Higgsing, the Hall--Littlewood index coincides with the Higgs branch Hilbert series \cite{Gadde:2011uv,Beem:2017ooy}.
The Hilbert series is a formal power series \cite{Gray:2008yu, Benvenuti:2010pq}
\begin{equation}
    \HS(t)=\sum_{n} c_nt^{n}\;,
\end{equation}
where the coefficients $c_n\in\N$ count the number of gauge invariant operators with scaling dimension $n$ constructed out of the scalar components of the hypermultiplets. Therefore, the Maclaurin series expansion of the index evaluated at the saddle should reproduce the number of gauge invariant operators of a given scaling dimension as the coefficients of the expansion. In order to extract the chiral ring generators, one can take the plethystic logarithm of the Hilbert series, which will be of the form
\begin{equation}
    \mathrm{PL}\left[\HS(t)\right]=\sum_{n}\tilde{c}_nt^n\;,
\end{equation}
where now the coefficients $\tilde{c}_n\in\mathbb{Z}$ count only chiral ring generators with positive sign, while chiral ring relations contribute with negative sign to the coefficients $\tilde{c}_n$.

The single-letter indices in the Hall--Littlewood specialisation are \cite{Gadde:2010en}
\begin{equation}\label{eq:HLsli}
    \sli_V=-t^2\;,\qquad \sli_H=t\;.
\end{equation}
Plugging these relations into \eqref{eq:deftnHL} we find $t_n=t^n$ for 4d $\mN=2$ theories, whereby the last term in \eqref{eq:IlongQ} drops out to leading order in $L$.
Thus the plethystic logarithm is simply obtained by discarding the summation over $n$ in \eqref{eq:IlongQ}, yielding 
\begin{equation}
    \left.\mathrm{PL}[\mI_{\mathfrak{su}} ]\right\rvert_{\substack{N \to \infty \\ L \to \infty}}=LN^2t\int\dd z\dd z'\varphi(z)G_1(z,z')\varphi(z') .
\end{equation}
We have also discarded the trace subtraction term, because it contributes to subleading order.

It is instructive to look at the small $t$ expansion of this expression. Using the derivative expansion of the Green's function
\begin{equation}
    G_1(z,z')=\sum_{k\geq 0}\left(\frac{t}{(1-t)^2}\right)^{k+1}\frac{1}{L^{2k}}\partial_z{}^{2k}\delta(z-z')\;,
\end{equation}
to the first few orders one finds
\begin{equation}
\begin{aligned}
   \left.\mathrm{PL}[\mI_{\mathfrak{su}} ]\right\rvert_{\substack{N \to \infty \\ L \to \infty}}=&LN^2\left[\int\dd z\varphi(z)^2 t^2+\int\dd z\left(2 \varphi(z)^2+\frac{1}{L^2}\varphi(z)\ddot{\varphi}(z)\right)t^3 \right.\\
    +&\left. \int\dd z\left(3\varphi(z)^2+\frac{4}{L^2}\varphi(z)\ddot{\varphi}(z)+\frac{1}{L^4}\varphi(z)\partial_z{}^4\varphi(z)\right)t^4\right]+\mathcal{O}\left(t^5\right)\;.
\end{aligned}\label{eq:PL series expansion}
\end{equation}\par

Our goal is to count gauge invariant operators constructed by contracting gauge indices of the scalar components of the hypermultiplets. Let us decompose the hypermultiplets as a pair of $\mathcal{N}=1$ chiral multiplets $(Q_{\alpha,i},\tilde{Q}_{i}^{\alpha})$, $\alpha=1,\dots, F_i$. Similarly, the bifundamental hypermultiplets $i \to i+1$ will be denoted as $(X_i,\tilde{X}_i)$. The lowest dimension gauge invariants are the mesons $Q_{\alpha,i}\tilde{Q}_i^{\beta}$, where there is no sum over the indices $\alpha,\beta,i$, and gauge index contraction is understood. There are $F_i{}^2$ such operators for each $i$. Therefore, one has in total $\sum_i F_i{}^2$ mesons, whose scaling dimension is $\Delta_{Q\tilde{Q}}=2$. This is clearly accounting for the $\mathcal{O}(t^2)$ coefficient of the continuum expression since
\begin{equation}
    \sum_i F_i^2\to LN^2\int\dd z\varphi(z)^2\;.
\end{equation}
The next lowest dimension operators are gauge invariants built out of 3 consecutive hypermultiplets, which should appear at $\mathcal{O}(t^3)$
\begin{equation}
    Q_{\alpha,i} X_i \tilde{Q}_{i+1}^{\beta}\;,\quad \tilde{Q}_i^{\alpha}\tilde{X}_i Q_{\beta,i+1}\;.
\end{equation}
There are exactly $2F_i F_{i+1}$ such operators for each $i$, and so taking the continuum limit, one recovers
\begin{equation}
   2 \sum_i F_i F_{i+1}=\sum_i F_i\left(F_{i+1}+F_{i-1}\right)\to LN^2\int\dd z\left(2\varphi(z)^2+\varphi(z)\frac{1}{L^2}\partial_z{}^2\varphi(z)\right)\;,
\end{equation}
which agrees with the $\mathcal{O}(t^3)$ coefficient of \eqref{eq:PL series expansion}. Similarly, at $\mathcal{O}(t^4)$ one expects the following `mesons' of length 4 for each fixed index
\begin{equation}
    Q_{\alpha,i} X_iX_{i+1} \tilde{Q}_{i+2}^{\beta}\;,\quad \tilde{Q}_i^{\alpha}\tilde{X}_i\tilde{X}_{i+1} Q_{\beta,i+2}\;,\quad Q_{\alpha,i}X_i\Tilde{X}_i\Tilde{Q}_i^\alpha\;,\quad 
\end{equation}
The first two mesons above constitute exactly $2F_iF_{i+2}$ gauge invariants for a fixed $i$, while the last operator contributes $F_i^2$ additional states totalling
\begin{equation}
    \sum_i \left(2F_iF_{i+2}+F_i^2\right)=\sum_iF_i(F_{i+2}+F_i+F_{i-2})\;.
\end{equation}
Let us add and subtract $-4F_{i+1}+6F_i-4F_{i-1}$ from the expression inside the bracket. Rearranging the resulting expression and taking the continuum limit as $L\to\infty$ one finds precise agreement with the $\mathcal{O}(t^4)$ coefficient of \eqref{eq:PL series expansion}.\footnote{We are using a symmetric definition for the continuum limit of the finite difference, analogous to \eqref{eq:continuum limit difference}.}

\subsection{Balanced quivers}
\label{sec:longbalanced}
An important class of quiver gauge theories, consists of those whose framings satisfy the balancing condition
\begin{equation}
    F_i=2N_i-N_{i+1}-N_{i-1}\;.
\end{equation}
In 4d, this is the condition for the vanishing of the beta-function, and consequently 4d holographic CFTs must necessarily satisfy this condition. In the long quiver limit, the balancing condition is translated to a differential constraint for the flavour rank function
\begin{equation}\label{eq:rankfunctionbalanced}
    \varphi(z)=-\frac{1}{L^2}\partial_z{}^2\nu(z)
\end{equation}
in terms of the rank function $\nu (z)$ with $\nu \left(\frac{j}{L}\right)=\frac{N_j}{N}$.
Plugging \eqref{eq:rankfunctionbalanced} into \eqref{eq:SeffSPELongQ}, and integrating by parts twice leads to
\begin{equation}
\begin{aligned}
\left.\mathrm{PL}[\mI_{\mathfrak{su}} ]\right\rvert_{\substack{N \to \infty \\ L \to \infty}}&=LN^2i_H(\mathbf{p})\int\dd z\nu(z)\varphi(z)\\
&+LN^2\int \dd z \dd z'\nu(z)\varphi(z')(i_V(\mathbf{p})+2i_H(\mathbf{p})-1)G_1(z,z') \;,
\end{aligned}
\end{equation}
where the boundary terms vanish automatically since we are interested in either periodic or Dirichlet boundary conditions for the Green's function.
Now consider the Hall--Littlewood limit of this index, 
\begin{equation}\label{eq:PLogbalanced1}
\begin{aligned}
    \left.\mathrm{PL}[\mI_{\mathfrak{su}} ]\right\rvert_{\substack{N \to \infty \\ L \to \infty}}&=\left(LN^2 \int\varphi(z)\nu(z)\dd z\right) t
    -\left(LN^2\int \dd z\dd z'\nu(z)\varphi(z')G_1(z,z')\right)(1-t)^2 .
\end{aligned}
\end{equation}
The significance of this expression is as follows. It is known that the order of the pole at $t\to 1$ in the Hilbert series gives the complex dimension of the Higgs branch \cite{Gray:2008yu}.

For the balanced quivers, the complex dimension of the Higgs branch is obtained by counting the difference of the number of hypermultiplets and vector multiplets \cite{Akhond:2021ffz}, 
\begin{equation}\label{eq:BalancedHBdim}
    2\sum_i\left[ N_i F_i+ N_i(N_{i+1}-N_i)\right]\to LN^2\int \nu(z)\varphi(z)\dd z \;.
\end{equation}
On the other hand, we want to study the asymptotics $t\to 1$ of the plethystic exponential of \eqref{eq:PLogbalanced1}, which we write as 
\begin{equation}\label{eq:PLbalancexpanded}
    \left.\mI_{\mathfrak{su}} \right\rvert_{\substack{N \to \infty \\ L \to \infty}} = \frac{1}{(1-t)^{LN^2\int\nu(z)\varphi(z)\dd z }} \PE\left[ g (t) \right]\;,
\end{equation}
where $g (t)$ is the second term in \eqref{eq:PLogbalanced1}. We show in appendix \ref{app:balancedt} that $\PE\left[ g (t) \right]$ is finite as $t \to 1$.
Thus the pole order of the Hall--Littlewood index \eqref{eq:PLbalancexpanded} matches \eqref{eq:BalancedHBdim}, constituting a highly non-trivial consistency check of our long quiver index.

\section{Applications}\label{sec:applications}
This section contains several explicit examples, showcasing our general results for the large $N$ index. 

\paragraph{}Our conventions for 4d $\mN=2$ indices follow \cite{Gadde:2020yah}. Concretely, letting $j_1,j_2$ be the Cartan generators of Spin$(4)$ and $R,r$ those of the R-symmetry $SU(2)_R\times U(1)_r$, and collectively denoting by $H$ the Cartan generators of the flavour symmetry, we define
\begin{equation}
    \mI^{\text{4d }\mN=2} = \tr_{\{\mathcal{Q},\mathcal{Q}^{\dagger}\}=0} (-1)^F p^{j_+ - \frac{1}{2} r_1}  q^{j_- - \frac{1}{2} r_1} u^{2 r_2} \mathbf{y}^H ,
\end{equation}
where $j_{\pm}=j_1\pm j_2$, $r_1=\frac{2}{3} \left( -2R + \frac{1}{2}\right)$ and $r_2=R+\frac{1}{2}$. Then, $\sli_{V}, \sli_{H}$ are explicitly given by \cite[Eq.(3.6)-(3.7)]{Gadde:2020yah}
\begin{equation}\label{eq:4dN2sli}
    \text{4d : } \quad \sli_V (\mathbf{p})= \frac{(pq)^{\frac{1}{3}} u^{-2} - (pq)^{\frac{2}{3}} u^{2}  -p-q+2pq}{(1-p)(1-q)} , \qquad \sli_H (\mathbf{p})= \frac{(pq)^{\frac{1}{3}} u - (pq)^{\frac{2}{3}} u^{-1} }{(1-p)(1-q)} .
\end{equation}\par
In 5d $\mN=1$, in our conventions the index is \cite{Kim:2012gu}
\begin{equation}
    \mI^{\text{5d }\mN=1} = \tr_{\{\mathcal{Q},\mathcal{Q}^{\dagger}\}=0} (-1)^F p^{j_+ +R}  q^{j_- +R} \mathbf{u}^{J} \mathbf{y}^H.
\end{equation}
In this case the R-symmetry is $SU(2)_R$, but in addition there is a global symmetry acting on instantons, whose Cartan generators we denoted by $J$. The perturbative index, which is expected to capture the leading order contribution at large $N$ in many cases, is $\left. \mI^{\text{5d }\mN=1}\right\rvert_{\mathbf{u}= 0}$. The single-letter indices are
\begin{equation}\label{eq:5dN1sli}
    \text{5d : } \quad \sli_V (\mathbf{p})= -\frac{p+q}{(1-p)(1-q)} , \qquad \sli_H (\mathbf{p})= \frac{(pq)^{\frac{1}{2}}}{(1-p)(1-q)} .
\end{equation}\par
In \eqref{eq:deftnHL} we have $t_n=t^n$, with 
\begin{equation}\label{eq:texplicit}
    \text{4d : } t = (pq)^{\frac{1}{3}}u , \qquad \text{5d : } t = (pq)^{\frac{1}{2}} .
\end{equation}

\subsection{4d affine exceptional quivers}
We start by providing the large $N$ index of affine $\hat{E}_{L-1}$ exceptional Dynkin quivers with gauge group $SU(N)^{L}$, where $L=7,8,9$ respectively. We spell out the results for 4d $\mN=2$ theories for definiteness, the 5d result being obtained in a completely analogous way. In absence of hypermultiplets other than the bifundamental, we apply \eqref{eq:ISUnofund} and compute
    \begin{equation}
    \begin{aligned}
        \left. \mI_{\mathfrak{su},\hat{E}_6} \right\rvert_{N\to\infty} &= \PE \left[ -7\sli_V (p,q,u) \right] \prod_{n=1}^{\infty} \frac{(1-p^n)^7(1-q^n)^7 (1-t^{2n})}{\left( 1-t^{4n}\right)\left( 1-t^{6n}\right)\left( 1-(tu^{-3})^n\right)^7}\\
        \left. \mI_{\mathfrak{su},\hat{E}_7} \right\rvert_{N\to\infty} &= \PE \left[ -8\sli_V (p,q,u) \right] \prod_{n=1}^{\infty} \frac{(1-p^n)^8(1-q^n)^8 (1-t^{2n})}{\left( 1-t^{4n}\right)\left( 1-t^{6n}\right)\left( 1-t^{8n}\right)\left( 1-(tu^{-3})^n\right)^8}\\
        \left. \mI_{\mathfrak{su},\hat{E}_8} \right\rvert_{N\to\infty} &= \PE \left[ -9\sli_V (p,q,u) \right] \prod_{n=1}^{\infty} \frac{(1-p^n)^9(1-q^n)^9 (1-t^{2n})^3}{\left( 1-t^{6n}\right)\left( 1-t^{8n}\right)\left( 1-t^{10n}\right)\left( 1-(tu^{-3})^n\right)^9} .
    \end{aligned}
    \end{equation}
For compactness we use the auxiliary $t$, which is a function of the fugacities given in \eqref{eq:texplicit}.\par
We also provide one orthosymplectic example, which we take for concreteness to be the higher-rank version of \cite[Eq.(3.8)]{Bourget:2020xdz}: this is the $\hat{E}_7$ Dynkin quiver with alternating $SO(2N_i)$ and $USp(2N_i)$ gauge nodes, in such a way that the central node is $USp(4N)$ and the three outermost nodes are $SO(2N)$. In this case, applying \eqref{eq:ISOSpuniversal}, the index can be written as:
\begin{equation}
    \left. \mI_{\mathfrak{osp}, \hat{E}_7}\right\rvert_{N\to\infty} = \prod_{n=1}^{\infty}\frac{\exp\left[ \frac{1}{n} \frac{(1+p^n)(1+q^n)}{(1-p^n)(1-q^n)}\frac{\left( 1-(tu^{-3})^n\right)\left(2-3 t^{2n}-12 t^{4n}-10 t^{6n}-12 t^{8n}-3 t^{10n}+2 t^{12n}\right)}{\left( 1+(tu^{-3})^n\right)\left(1+t^{4n}\right)\left(1-14t^{4n}+t^{8n}\right)}\right] }{\sqrt{\left(\frac{1-(tu^{-3})^{n}}{(1-p^n)(1-q^n)}\right)^8 \left(1-t^{4n}\right)^2\left(1-6t^{2n}+t^{4n}\right) \left(1-14t^{2n}+t^{4n}\right) }} 
\end{equation}
In the Hall--Littlewood limit, we take the plethystic logarithm and obtain 
\begin{equation}
    \left. \mathrm{PL}\left[\mI_{\mathfrak{osp}, \hat{E}_7}\right] \right\rvert_{N\to\infty} =7 t^2 + 77 t^4 + 441 t^6 + 5079 t^8 + 52419 t^{10} + \mathcal{O}(t^{12}) .
\end{equation}

\subsection{5d \texorpdfstring{$E_{N_f+1}$}{ENf} theories}

We apply our orthosymplectic formula \eqref{eq:ISOSpuniversal} to compute the superconformal index of the 5d $\mN=1$ higher-rank $E_{N_f+1}$ theories studied holographically in \cite{Bergman:2012kr}. These are $USp(2N)$ gauge theories with one antisymmetric hypermultiplet (counted as $\Lambda=2$ half-hypers), plus $0 \le N_f \le 7$ fundamental hypermultiplets. From \eqref{eq:SOSpIndexvi} we have in this case
\begin{equation}
\begin{aligned}
    v_{2m}& = \sli_{H}(\mathbf{p}^{2m}) \chi_{N_f} (\mathbf{y}^{2m})  + \sli_{V}(\mathbf{p}^{m}) - 2\sli_{H}(\mathbf{p}^{m}) , \qquad v_{2m+1}= \sli_{H}(\mathbf{p}^{2m+1}) \chi_{N_f} (\mathbf{y}^{2m+1}) ,
\end{aligned}
\end{equation}
and $h_n =  \sli_{V}(\mathbf{p}^{n}) + 2\sli_{H}(\mathbf{p}^{n})$. The index is given by 
\begin{equation}\label{eq:ISOSpENf}
\begin{aligned}
    \left.\mI_{\text{5d }E_{N_f+1}} \right\rvert_{N\to\infty} &= \exp \left\{\frac{1}{2}\sum_{m=1}^{\infty} \left[ \frac{1}{2m}\frac{\left( \sli_{H}(\mathbf{p}^{2m}) \chi_{N_f} (\mathbf{y}^{2m})  + \sli_{V}(\mathbf{p}^{m}) - 2\sli_{H}(\mathbf{p}^{m}) -1\right) ^2}{1-\sli_{V}(\mathbf{p}^{2m}) - 2\sli_{H}(\mathbf{p}^{2m}) } \right. \right.\\
    & \left. \left.+ \frac{1}{2m-1}\frac{\left( \sli_{H}(\mathbf{p}^{2m-1}) \chi_{N_f} (\mathbf{y}^{2m-1})\right) ^2}{1-\sli_{V}(\mathbf{p}^{2m-1}) - 2\sli_{H}(\mathbf{p}^{2m-1}) } \right] \right\}  \prod_{n=1}^{\infty}\left( 1-\sli_{V}(\mathbf{p}^{n}) - 2\sli_{H}(\mathbf{p}^{n})\right)^{-\frac{1}{2}} .
\end{aligned}
\end{equation}
In particular, the case $N_f=0$ reproduces the result of \cite{Bergman:2013koa}. In the unrefined limit of \eqref{eq:ISOSpENf}, stripping off the overall factor $(1-t)^2$ due to the reducibility of the antisymmetric representation (cf. Remark \ref{rmk:irrepPESOSp}), and taking the plethystic logarithm we get 
\begin{equation}
     \left. \mathrm{PL}\left[\mI_{\text{5d }E_{N_f+1}}\right] \right\rvert_{N\to\infty} =  \left( 3 + N_f(2N_f-1)\right) t^2 + \left( 4+ 2N_f (2N_f-1)\right) t^3 + \mathcal{O}(t^4) .
\end{equation}
The coefficient of $t^2$ equals $\dim \mathfrak{usp}(2)+ \dim \mathfrak{so}(2N_f)$, matching the dimension of the flavour symmetry. Counting gauge invariant operators with scaling dimension 3, we have $4$ gauge invariant tensors formed by the symmetric product of three antisymmetric half-hypermultiplets, plus $2\dim \mathfrak{so}(2N_f)=2N_f (2N_f-1)$ gauge invariant generators formed by any 2 of the fundamentals paired up with either of the antisymmetric half-hypermultiplets; this matches the coefficient of $t^3$.\par

\subsection{Hilbert series of Hitchin moduli spaces}
\label{sec:Hitchin}

\begin{figure}[ht]
    \centering
    \begin{tikzpicture}
        \node (t1) at (-2,1) {$T_\Sigma$};
        \node (t2) at (2,1) {$T^{\vee}_{\text{4d}}$};
        \node (t3a) at (-2,-1) {$T_{\text{3d}}$};
        \node (t3b) at (2,-1) {$T^{\vee}_{\text{3d}}$};

        \path[->] (t1) edge node[anchor=east] {\footnotesize $S^1$} (t3a);
        \path[->] (t2) edge node[anchor=west] {\footnotesize $S^1$} (t3b);
        \path[->] (t3a) edge node[anchor=south] {\footnotesize mirror} (t3b);

        \node (eq) at (8,0) {HB$(T^{\vee}_{\text{4d}})\underset{S^1}{\cong}\ $HB$(T^{\vee}_{\text{3d}})\underset{\text{mirror}}{\cong}$CB$(T_{\text{3d}})\cong \mathcal{M}_{\Sigma}$};
    \end{tikzpicture}
    \caption{Left: The bottom-right theory $T^{\vee}_{\text{3d}}$ is the 3d mirror to the top-left theory $T_\Sigma$. When $T^{\vee}_{\text{3d}}$ is Lagrangian, we can construct a 4d Lagrangian theory $T^{\vee}_{\text{4d}}$ with the same Higgs branch. The vertical arrows represent $S^1$ compactifications, and the horizontal arrow represents 3d mirror symmetry. Right: The Higgs branch of $T^{\vee}_{\text{4d}}$ equals by construction the Coulomb branch of $T_{\text{3d}}$, which is a moduli space of solutions to Hitchin's system associated to $\Sigma$.}
    \label{fig:3d4dmirror}
\end{figure}

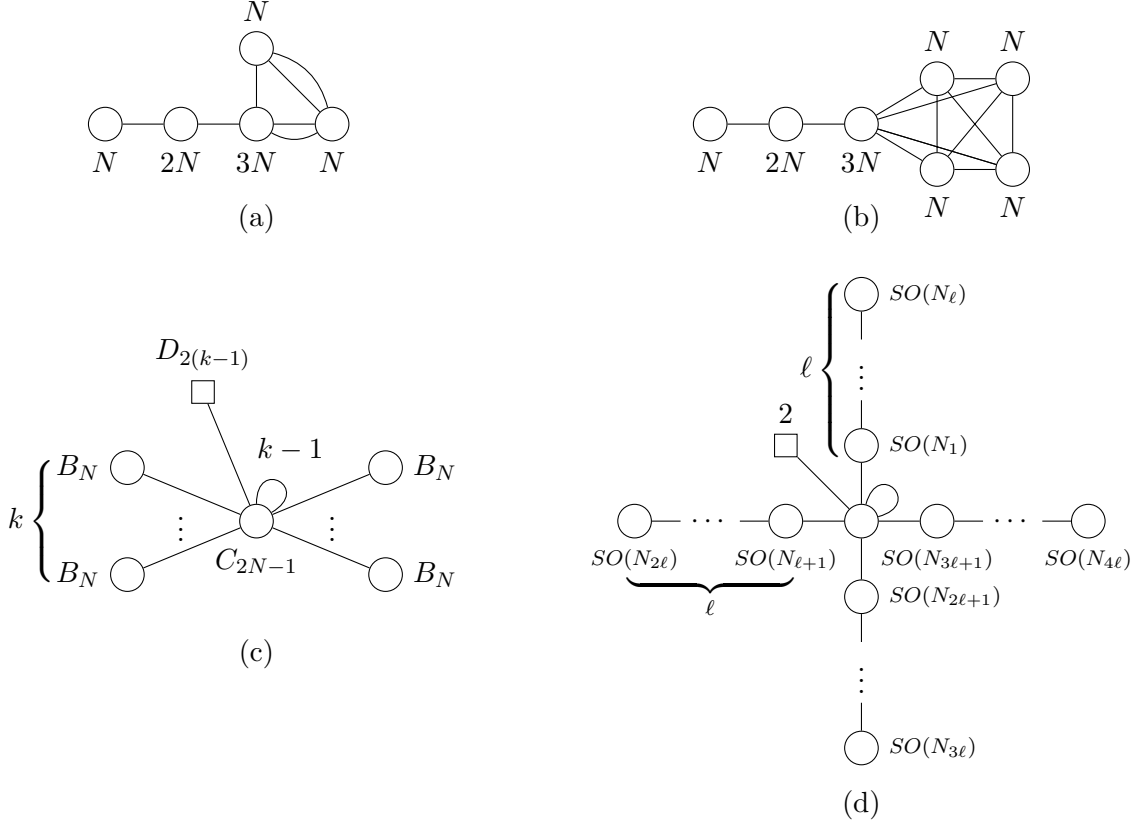
\begin{figure}[ht]
    \centering
    \begin{tikzpicture}
        \node[label=below:{$3N$}][u](N3) at (-4,2) { \hspace{8pt} };
        \node[label=below:{$2N$}][u](N2)[left of=N3]{ \hspace{8pt} };
        \node[label=below:{$N$}][u](N1)[left of=N2] { \hspace{8pt} };
        \node[label=below:{$N$}][u](N4)[right of=N3]{ \hspace{8pt} };
        \node[label=above:{$N$}][u](N5)[above of=N3]{ \hspace{8pt} };

        \draw (N1) -- (N2) -- (N3) -- (N4) -- (N5) -- (N3);
        \path (N4) edge[bend left] (N3) edge[bend right] (N5);

        \node[label=below:{$3N$}][u](M3) at (4,2) { \hspace{8pt} };
        \node[label=below:{$2N$}][u](M2)[left of=M3]{ \hspace{8pt} };
        \node[label=below:{$N$}][u](M1)[left of=M2] { \hspace{8pt} };
        \node[label=below:{$N$}][u](M4a) at (5,1.4) { \hspace{8pt} };
        \node[label=above:{$N$}][u](M5a) at (5,2.6) { \hspace{8pt} };
        \node[label=below:{$N$}][u](M4b)[right of=M4a]{ \hspace{8pt} };
        \node[label=above:{$N$}][u](M5b)[right of=M5a]{ \hspace{8pt} };
        \draw (M1) -- (M2) -- (M3) -- (M4a) -- (M5a) -- (M5b) -- (M4b) -- (M3) -- (M5a);
        \draw (M5a) -- (M4b) -- (M4a) -- (M5b);
        \draw (M3) -- (M5b);
        \draw (M3) -- (M4b);

        \node[label=below:{$ C_{2N-1}$}][u](Sp) at (-4,-3.25) { \hspace{8pt} };
        \node (O1) [left of=Sp]{$\vdots$};
        \node (O2)[right of=Sp]{$\vdots$};
        \node[label=left:{$B_N$}][u](O1b)[above left of=O1]{ \hspace{8pt} };
        \node[label=left:{$B_N$}][u](O1c)[below left of=O1]{ \hspace{8pt} };
        \node[label=right:{$B_N$}][u](O2b)[above right of=O2]{ \hspace{8pt} };
        \node[label=right:{$B_N$}][u](O2c)[below right of=O2]{ \hspace{8pt} };
        \node (empty) [above of=Sp]{};
        \node[label=above:{$D_{2(k-1)}$}][uf](F1)[above left of=empty]{ \hspace{12pt}};
        \node[anchor=west] at (-4.125,-2.3) {$ k-1$};
        \node at (-6.7,-3.25) {$k \begin{cases} \ \\ \ \\ \ \\  \end{cases}$};

        \draw (F1) -- (Sp);
        
        \draw (O1b) -- (Sp);
        \draw (O1c) -- (Sp);
        \draw (O2b) -- (Sp);
        \draw (O2c) -- (Sp);
        \loopedge{Sp}{90}{30};

        \node[u](Sp) at (4,-3.25) { \hspace{8pt} };
        \node[label=right:{$\scriptstyle SO(N_1)$}][u](On1)[above of=Sp]{ \hspace{8pt} };
        \node (Ond) [above of=On1]{$\vdots$};
        \node[label=right:{$\scriptstyle SO(N_{\ell})$}][u](On2)[above of=Ond]{ \hspace{8pt} };
        \node[label=below:{$\scriptstyle SO(N_{\ell+1})$}][u](Ow1)[left of=Sp]{ \hspace{8pt} };
        \node (Owd) [left of=Ow1]{$\cdots$};
        \node[label=below:{$\scriptstyle SO(N_{2\ell})$}][u](Ow2)[left of=Owd]{ \hspace{8pt} };
        \node[label=below:{$\scriptstyle SO(N_{3\ell+1})$}][u](Oe1)[right of=Sp]{ \hspace{8pt} };
        \node (Oed) [right of=Oe1]{$\cdots$};
        \node[label=below:{$\scriptstyle SO(N_{4\ell})$}][u](Oe2)[right of=Oed]{ \hspace{8pt} };
        \node[label=right:{$\scriptstyle SO(N_{2\ell+1})$}][u](Os1)[below of=Sp]{ \hspace{8pt} };
        \node (Osd) [below of=Os1]{$\vdots$};
        \node[label=right:{$\scriptstyle SO(N_{3\ell})$}][u](Os2)[below of=Osd]{ \hspace{8pt} };
        
        \node[label=above:{$2$}][uf](F1) at (3,-2.25) { \hspace{12pt}};
        \node at (3.75,-1.25) {$\ell \begin{cases} \ \\ \ \\ \ \\ \ \\ \end{cases}$};
        \node at (2,-4.25) {$\underbrace{\hspace{2.2cm}}_{\ell}$};

        \draw (F1) -- (Sp);        
        \draw (On2) -- (Ond) -- (On1) -- (Sp);
        \draw (Ow2) -- (Owd) -- (Ow1) -- (Sp);
        \draw (Oe2) -- (Oed) -- (Oe1) -- (Sp);
        \draw (Os2) -- (Osd) -- (Os1) -- (Sp);
        \loopedge{Sp}{15}{75};

        \node at (-4,.75) {(a)};
        \node at (4,.75) {(b)};
        \node at (-4,-5) {(c)};
        \node at (4,-7) {(d)};

    \end{tikzpicture}
    \caption{A sample of 4d quiver theories $T^{\vee}_{\mathrm{4d}}$. Unitary groups in (a) and (b) are indicated by denoting their ranks, while orthogonal, or symplectic groups in (c) are indicated by their associated algebras. The loop attached to the $C_{2N-1}$ gauge node in (c) denotes the presence of $(k-1)$ hypermultiplets in the second rank antisymmetric representation. (d) consists of 4 orthosymplectic quiver tails of length $\ell$ attached to a central $USp(4N-2)$ node, plus 2 fundamental and 1 antisymmetric hypermultiplets.}
    \label{fig:HitchinMirrorQ}
\end{figure}

We apply our formulas \eqref{eq:Iuniversal}-\eqref{eq:ISOSpuniversal} to study the leading large $N$ behaviour of the ring of functions on Hitchin moduli spaces. 
We use the strategy summarised in figure \ref{fig:3d4dmirror}, leveraging the 3d mirror of theories of class $S$ \cite{Benini:2010uu,Xie:2012hs}.
Our aim is by no means to be exhaustive, but simply to showcase one further application of our results in a handful of selected examples, listed in figure \ref{fig:HitchinMirrorQ}.\par

\begin{enumerate}[(i)]
    \item Given a 4d $\mN=2$ theory $T_{\Sigma}$ of class $S$, typically non-Lagrangian, we can compactify it on a circle and obtain an effective 3d $\mN=4$ theory $T_{\text{3d}}$. The Coulomb branch of $T_{\text{3d}}$ is a moduli space $\mathcal{M}_{\Sigma}$ parametrising solutions to Hitchin's system associated to the punctured Riemann surface $\Sigma$ \cite{Kapustin:1998xn} (or more precisely, a scaling limit thereof).
    \item A 3d mirror $T^{\vee}_{\text{3d}}$ to $T_{\text{3d}}$ will have, by construction, a Higgs branch isomorphic to $\mathcal{M}_{\Sigma}$, in one complex structure. We restrict our attention to $T_{\Sigma}$ such that $T^{\vee}_{\text{3d}}$ has a Lagrangian description. In this case, there exists a 4d $\mN=2$ quiver gauge theory $T^{\vee}_{\text{4d}}$ whose Higgs branch, upon circle compactification, matches the Higgs branch of $T^{\vee}_{\text{3d}}$.
    \item We can immediately obtain the large $N$ Hall--Littlewood index of $T^{\vee}_{\text{4d}}$ as a specialisation of the general formulas \eqref{eq:Iuniversal}-\eqref{eq:ISOSpuniversal}. This equals, by construction, the leading large $N$ contribution to the Hilbert series of $\mathcal{M}_{\Sigma}$. This gives an upper bound on the counting of generators of the ring of functions on $\mathcal{M}_{\Sigma}$.
\end{enumerate}
While we do not explore here the behaviour of the moduli spaces $\mathcal{M}_{\Sigma}$ at large $N$, we provide a sample of large $N$ Hall--Littlewood indices computed following the circle of ideas just outlined.\par

\par

All the Lagrangian theories in figure \ref{fig:HitchinMirrorQ} satisfy the complete Higgsing condition \cite{Beem:2017ooy}, hence their Hall--Littlewood index equals the Higgs branch Hilbert series.
Quivers (a) and (b) are higher-rank versions of examples in \cite[Fig.17]{Xie:2012hs}; in the notation of \cite{Xie:2012hs}, they are realised by patterns of Young tableaux labelling the irregular singularities:
\begin{equation}\label{eq:Xietableaux}
\begin{aligned}
    \text{(a) : }& [3N, N] \to [3N, N, N ] \to [3N, N, N ]  \to [N, N,N, N, N ] , \\ 
    \text{(b) : }& [N, N,N, N] \to [N, N,N, N] \to [N, N,N, N] \quad + \text{ full regular puncture} .
\end{aligned}
\end{equation}
These quivers have no framing, thus their indices are straightforwardly obtained applying \eqref{eq:ISUnofund}:
\begin{align}
    \left.\mI_{\text{(a)}}\right.\rvert_{N\to\infty} &= \prod_{n=1}^{\infty} \frac{1}{\left(t^{2 n}-t^n+1\right) \left(t^{2 n}+t^n+1\right) \left(-7 t^{2 n}-8 t^{3 n}-7 t^{4
   n}+t^{6 n}+1\right)} \\
   \left.\mI_{\text{(b)}}\right.\rvert_{N\to\infty} &= \prod_{n=1}^{\infty} \frac{1}{\left(1+t^n+t^{2 n}\right)^3 \left(1-3 t^n-2 t^{2 n}-3 t^{3 n}-2 t^{4 n}-3 t^{5 n}-2 t^{6 n}-3
   t^{7 n}+t^{8 n}\right)}  .
\end{align}
\par
\medskip
Quiver (c) is a family of examples from \cite{Kang:2022zsl}, labelled by $k\in \mathbb{N}$. It is orthosymplectic and contains fundamental flavours, so our results become especially useful. Quiver (c) consists of a central node with gauge group $USp(4N-2)$, attached to $2k$ nodes with gauge groups $SO(2N+1)$, together with $(k-1)$ antisymmetric hypermultiplets and $2(k-1)$ fundamental hypermultiplets \cite{Kang:2022zsl}. We apply \eqref{eq:ISOSpuniversal} and, defining for shortness the functions 
\begin{equation}
\begin{aligned}
    f_{\text{(c)}}^{\text{even}} (t) &= \frac{\left(1+t^2\right) \left(1+2 (k-1)t^{\frac{1}{2}} -(6 k-5) t-\frac{2 k (1-t)t^2}{1+t^2}\right)^2}{1-2 (k-1) \left(t+t^2+t^3\right)+t^4}+\frac{2 k (1-t)^2t^2}{1+t^2} -1 \\
    f_{\text{(c)}}^{\text{odd}} (t) &=\frac{16 (k-1)^2 t^2 \left(1+t^2\right)}{1-2 (k-1) \left(t+t^2+t^3\right)+t^4}+2k t^2 ,
\end{aligned}
\end{equation}
we get:
\begin{align}
    \left.\mI_{\text{(c)}}\right.\rvert_{N\to\infty} &= \prod_{n=1}^{\infty} \sqrt{\frac{\exp \left[ \frac{1}{n}\left( \delta_{n \in 2 \Z} f_{\text{(c)}}^{\text{even}} (t^n) + \delta_{n+1 \in 2 \Z} f_{\text{(c)}}^{\text{odd}} (t^n)  \right)\right]}{\left(1+t^{2 n}\right)^{2k-1} \left[1-2(k-1) (t^n+ t^{2 n}+ t^{3 n})+t^{4 n}\right]}} .
\end{align}
Stripping off the factor $(1-t)^{2k}$ from the free hypermultiplets and taking the plethystic logarithm we obtain 
\begin{equation}\label{eq:PLogExc}
\begin{aligned}
    \left.\mathrm{PL}\left[(1-t)^{2k} \mI_{\text{(c)}}\right]\right.\rvert_{N\to\infty} &=  (k-1)(10k-11) t^2 + \frac{2}{3}(k-1)(26k^2-55k+30) t^3 \\
    &+ \frac{1}{2} \left(40-181 k+299 k^2-228 k^3+68 k^4\right) t^4 \\
    &+\frac{2}{5} \left(-65+357k-840 k^2+1040 k^3-660 k^4+168 k^5\right) t^5 + \mathcal{O} (t^6) .
\end{aligned}
\end{equation}
We observe that 
\begin{equation}
    \dim \mathfrak{so}(4(k-1))+\dim \mathfrak{usp}(2(k-1)) = (k-1)(10k-11) ,
\end{equation}
so the coefficient of $t^2$ in \eqref{eq:PLogExc} matches the dimension of the flavour symmetry. We can form gauge invariant operators of dimension 3 in two ways: by pairing two half-hypermultiplets with any of the antisymmetric half-hypermultiplets, or forming cubic tensors by the symmetric product of three antisymmetric half-hypermultiplets. The two contributions sum up to
\begin{equation}
    \frac{(4k-4)(4k-5)}{2} \cdot 2 (k-1) + \left( \begin{matrix}3-1+2(k-1) \\ 3\end{matrix}\right) = 4(k-1)^2(4k-5)+\frac{2}{3} k (k-1) (2k-1)  ,
\end{equation}
with the sum reproducing the coefficient of $t^3$ in \eqref{eq:PLogExc}. 
\par
\medskip

Let us consider one more family of examples from \cite{Kang:2022zsl}. Here we set $k=2$ but allow for longer orthosymplectic tails, which we take to consist of $\ell$ nodes each. In total the quiver has $\lvert \ver \rvert=4\ell +1$, and there are 2 fundamental hypermultiplets and one antisymmetric hypermultiplet charged under the gauge group $USp(4N-2)$ at the central node. The gauge ranks of the legs are determined by partitions of $4N-2$ of length $\ell$, and for simplicity we assume $\ell$ is odd and that the ranks are $N_i =\mathcal{O}(N)$.\par 
Applying the Hall--Littlewood specialisation of \eqref{eq:ISOSpuniversal}, removing the contribution of the decoupled hypermultiplet and taking the plethystic logarithm, we arrive at:
\begin{equation}
\begin{tabular}{|c|c|}
\hline 
$\ell$ &  $\left.\mathrm{PL}\left[(1-t)^2 \mI_{\text{(d)}}\right]\right.\rvert_{N\to\infty}$ \\
\hline
1  & $9 t^2+16 t^3+69 t^4+170 t^5+521 t^6+1416 t^7+4174 t^8+11758 t^9+ \mathcal{O}(t^{10}) $\\
3  & $9 t^2+16 t^3+69 t^4+178 t^5+557 t^6+1584 t^7+4812 t^8+14130 t^9+ \mathcal{O}(t^{10}) $\\
5  & $9 t^2+16 t^3+69 t^4+178 t^5+557 t^6+1584 t^7+4812 t^8+14138 t^9+ \mathcal{O}(t^{10}) $\\
7  & $9 t^2+16 t^3+69 t^4+178 t^5+557 t^6+1584 t^7+4812 t^8+14138 t^9+ \mathcal{O}(t^{10})$\\
\hline
\end{tabular}
\end{equation}
The case $\ell=1$ reduces to the case $k=2$ of example (c). For every $\ell$, the coefficients of $t^2$ and $t^3$ are determined exactly as for $k=1$ above, because there are no other possible generators of dimensions 2 and 3, respectively. The coefficients of $t^n$ are necessarily independent of $\ell$ for $n\le 2\ell +2 $.

\subsection{Universality of Hagedorn temperature}
\label{sec:Hagedorn}

We now make the observation that the leading divergence of the thermal partition function near the Hagedorn temperature can be computed from the leading divergence of the superconformal index, if we write 
\begin{equation}\label{eq:identifyT}
    \Re \left(\log (q) \right),\Re \left(\log (p)\right)\propto -\frac{2\pi}{T}  
\end{equation}
and identify this parameter $T$ with the temperature of the  thermal partition function.\par

We thus define the Hagedorn temperature $T_H$ as the value of $T$ at which our solution develops a singularity. Concretely, this happens when one of the eigenvalues of a matrix $[\mathbb{I}-h_n]$ vanishes.
Inspecting our general results \eqref{eq:Iuniversal}-\eqref{eq:ISOSpuniversal}, we immediately deduce the following \emph{universal properties} of the Hagedorn temperature $T_H$ in quiver theories with 8 supercharges:
\begin{itemize}
    \item It is always independent of the number of fundamental hypermultiplets.
    \item $T_H$ will in general depend on the pattern of adjoint, symmetric and antisymmetric hypermultiplets charged under each gauge node. However, if there are the same numbers $g,S,\Lambda$ of such hypermultiplets $\forall i \in \ver$, $T_H$ depends on the details of the theory only through the largest eigenvalue $\lambda_{\mathrm{max}} (\mathbb{A})$ of the adjacency matrix $\mathbb{A}$ of the quiver, via
    \begin{equation}\label{eq:THdefeqSU}
         t^n+ t^{-n}  -\left(g + S + \Lambda\right) - \lambda_{\mathrm{max}} (\mathbb{A}) =0.
    \end{equation}
    For ADE quivers with $g=S=\Lambda=0$, this equation reduces to the result in \cite{Calderon-Infante:2026rkj}, providing a consistency check for our identification \eqref{eq:identifyT}; \eqref{eq:THdefeqSU} holds for arbitrary quivers. 
\end{itemize}
We thus have a comprehensive classification of the behaviour of the Hagedorn temperature, stemming from \eqref{eq:Iuniversal}-\eqref{eq:ISOSpuniversal} imposing the identification \eqref{eq:identifyT}. This extends the result of \cite{Calderon-Infante:2026rkj} to quivers of arbitrary shape, gauge groups, and matter content.\par
It was noted in \cite{Calderon-Infante:2026rkj} that all holographic quivers with special unitary gauge groups have the same Hagedorn temperature, corresponding to $\lambda_{\mathrm{max}} (\mathbb{A}) =2$ in \eqref{eq:THdefeqSU}. We extend the result: \emph{every} holographic quiver has the same Hagedorn temperature.
Notice that the converse is not true: we can produce infinite families of quiver gauge theories with the same $T_H$. 
These quivers, however, are over-balanced, thereby not holographic.\par
These corollaries answer many of the open questions listed in \cite{Calderon-Infante:2026rkj}.

\section{Outlook}
In this work we examined the superconformal index of theories with 8 supercharges in the large $N$, and long quiver limits. Our exact expression for the index was shown to be consistent with the Hall--Littlewood limit, as it correctly enumerates the spectrum of Higgs branch chiral ring generators.

We implemented the $SU(N)$ gauge theories by extracting the overall $U(1)$ contribution of the vector multiplet to leading order, but a treatment that keeps track of the finite-$N$ distinction between $U(N)$ and $SU(N)$ would be desirable. In particular, the chiral ring for the latter contains baryons, whose extraction from the Hilbert series at subleading orders in $N$ would be a natural extension of this work. 

It is natural to consider extending the results presented here to the 3d index, where the summation over monopoles makes the computation more technically involved. The methods provided here should nevertheless apply. One advantage of considering 3d theories, is that mirror symmetry provides further stringent consistency tests of the long quiver solution. The authors wish to report on this matter in a future publication.

A major motivation for the current work is to provide further observables for holography. Indeed, a large class of warped AdS vacua in type II string theory are conjectured to be holographically dual to long quivers \cite{Reid-Edwards:2010vpm, Aharony:2012tz,Nunez:2019gbg, Akhond:2021ffz, DHoker:2016ujz, DHoker:2017mds, DHoker:2017zwj}. The exact expressions presented in this work pave the way for future precision tests of these holographic duals. What complicates the task is the intricate, and model dependent nature of the internal space of the solutions of interest here. In the case of the AdS$_6$ solutions, however, the consistent truncations to Romans $F_4$ supergravity were obtained in \cite{Hong:2018amk}, which are likely to prove useful in the task of matching the index from the bulk.\par
Another intended application of our results is to the giant graviton expansion, which can be isolated by evaluating the finite $N$ index order by order in the fugacities and normalising by the large $N$ value. For 4d $\mN=1$ theories \emph{without} fundamental flavours, the large $N$ index of \cite{Gadde:2010en,Eager:2012hx} can be used to test the holographic dual giant gravitons \cite{GonzalezLezcano:2026wje} in the product of AdS$_5$ with a Sasaki--Einstein five-manifold; see also \cite{Purkayastha:2025whi}. Likewise, our formula \eqref{eq:Iuniversal} (see Remark \ref{rmk:4dN1index}) is useful for a holographic prediction for the giant graviton expansion in theories with unquenched flavours \cite{Benini:2006hh,Benini:2007gx}.

\bigskip
\paragraph{Acknowledgements.} The authors wish to thank Yuanyuan Fang, Ali Fatemiabhari, Noppadol Mekareeya, Carlos Nunez, Shlomo Razamat, Ricardo Stuardo, Miguel Tierz, Christoph Uhlemann, and Zhenghao Zhong for discussions related to this work. MA is supported in part by an INFN fellowship under the ``Iniziativa Specifica'' 24 ST\&FI.
\paragraph{AI usage.} Claude Opus 5 (Anthropic) was used to check the consistency conditions for \eqref{eq:Xietableaux}, which were then confirmed by the authors. It also provided some useful insight in appendix \ref{app:balancedt}, although its proof was flawed for $t \notin (0,1)$. GPT-5.6 Luna (OpenAI) was used for assistance with proofreading and improving the clarity of the manuscript.

\begin{appendix}
\section{General orthosymplectic quiver matrix models}
\label{app:SOSpderivation}
We now derive the large $N$ limit of orthosymplectic quiver matrix models.\footnote{In the single-node case, similar manipulations have appeared for instance in  \cite{Santilli:2020ueh,Santilli:2021eon}, and a different regime of the superconformal index of orthosymplectic single-node theories has been considered in \cite{Amariti:2020jyx,Amariti:2021ubd}.}
Consider a matrix integral 
\begin{equation}\label{eq:ZSOSp}
    \mI = \int_{G} \dd\mu_{G}(X) f (X) ,
\end{equation}
where $G$ is a product of special orthogonal and symplectic groups, as in \eqref{eq:SOSpgauge}, and $\dd\mu_{G}=\prod_{j \in \ver} \dd\mu_{G_j}$ is the normalised Haar measure on \eqref{eq:SOSpgauge}. We parametrise the maximal torus of each $G_j$ by the $N_j$ independent eigenvalues $e^{\ii \theta_{a,j}}$, with $\theta_{a,j} \in [0,\pi]$ for $a=1,\dots, N_j$. 
In terms of these eigenvalues, we have (see for instance \cite[\S2.6]{ForresterBook}):
\begin{equation}\label{HaarSOSp}
\begin{aligned}
	\dd \mu_{SO(2N_j)} &= \frac{1}{C_j}\prod_{1 \le a < b \le N_j} \left( 2 \sin \frac{\theta_{a,j}-\theta_{b,j}}{2} \right)^2\left( 2 \sin \frac{\theta_{a,j}+\theta_{b,j}}{2} \right)^2  \prod_{a=1}^{N_j} \frac{\dd \theta_{a,j}}{2\pi} \\
	\dd\mu_{SO(2N_j+1)} (\theta_j) &= \frac{1}{C_j}\prod_{1 \le a < b \le N_j} \left( 2 \sin \frac{\theta_{a,j}-\theta_{b,j}}{2} \right)^2\left( 2 \sin \frac{\theta_{a,j}+\theta_{b,j}}{2} \right)^2  \prod_{a=1}^{N_j} \left(2 \sin \frac{ \theta_{a,j}}{2}\right)^2 \frac{\dd \theta_{a,j}}{2\pi} \\
	\dd\mu_{USp(2N_j)} (\theta_j) &= \frac{1}{C_j}\prod_{1 \le a < b \le N_j} \left( 2 \sin \frac{\theta_{a,j}-\theta_{b,j}}{2} \right)^2\left( 2 \sin \frac{\theta_{a,j}+\theta_{b,j}}{2} \right)^2  \prod_{a=1}^{N_j} \left(2 \sin  \theta_{a,j}\right)^2 \frac{\dd \theta_{a,j}}{2\pi} ,
\end{aligned}
\end{equation}
where $C_j$ are normalisation constants. In \eqref{eq:ZSOSp}, $f$ is a class function of the generic form 
\begin{equation}\label{eq:fSOSp}
\begin{aligned}
    f(X)=&\exp \left\{ \sum_{n\ge1}\frac{1}{n}\left[ \sum_{i\in \ver}  \left( \sum_{a=1}^{N_i}\sum_{\epsilon_a = \pm 1} v^{i}_n \cos (n \epsilon_a \theta_{a,i}) \right. \right.\right.\\ 
    +& \left. \sum_{a=1}^{N_i}\sum_{\epsilon_a = \pm 1}\sum_{b=1}^{N_i}\sum_{\epsilon_b = \pm 1} \frac{h^{ii}_n}{2}\cos (n (\epsilon_a \theta_{a,i} - \epsilon_b \theta_{b,i}))  +  \sum_{a=1}^{N_i}\sum_{\epsilon_a = \pm 1} \kappa_i h^{ii}_n \cos (n \epsilon_a \theta_{a,i})\right)\\
    +& \sum_{\substack{ (i,j)\in \ver^2 \\ i < j }} \left( \sum_{a=1}^{N_i}\sum_{\epsilon_a = \pm 1} \sum_{b=1}^{N_j}\sum_{\epsilon_b = \pm 1} h^{ij}_n\cos (n (\epsilon_a \theta_{a,i} - \epsilon_b \theta_{b,j})) \right. \\
    +& \left.\left.\left. \kappa_i h^{ij}_n \sum_{b=1}^{N_j}\sum_{\epsilon_b = \pm 1}\cos (n \epsilon_b \theta_{b,j}) +  \kappa_j h^{ji}_n \sum_{a=1}^{N_i}\sum_{\epsilon_a = \pm 1}\cos (n \epsilon_a \theta_{a,i})  \right)  \ +c_n \right] \right\}.
\end{aligned}
\end{equation}
Here the coefficient $\kappa_j =1 $ if $G_j=SO(2N_j+1)$ and $0$ otherwise, as defined in \eqref{defshiftSOSp}. Besides, observe that the coefficients have different factors of 2 compared to the unitary case, because all representations involved in the orthosymplectic case are (pseudo-)real. Finally, the term $c_n$ collects all contributions from the groups $SO(2N_i+1)$ which are independent of the eigenvalues, thus are spectators throughout the derivation.\par
We write \eqref{eq:ZSOSp} in the form 
\begin{equation}
    \mI = \exp\left( \sum_{n\ge1} \frac{c_n}{n}\right) \prod_{j\in\ver}\prod_{c=1}^{N_j} \dashint_{0}^{\pi} \frac{\dd \theta_{c,j}}{2\pi} \exp \left( S_{\mathrm{eff}}^{\mathfrak{osp}} \right) ,
\end{equation}
where we have dropped an overall constant that eventually cancels against the normalisation, and factored out the last piece of \eqref{eq:fSOSp}. The effective action is given by 
\begin{equation}
\begin{aligned}
    S_{\mathrm{eff}}^{\mathfrak{osp}} = &\sum_{n\ge1}\frac{1}{n} \left\{ \sum_{i\in\ver} \sum_{a=1}^{N_i}\cos (n \theta_{a,i})  \left[ 2 v^{i}_n +2\kappa_i h^{ii}_n + 2\sum_{\substack{j\in \ver \\ j>i}} \kappa_j (h^{ij}_n + h^{ji}_n) \right] \right. \\
     & + \sum_{(i,j)\in\ver^2}  \sum_{a=1}^{N_i}\sum_{b=1}^{N_j} \sum_{\epsilon=\pm 1} \left[ h^{ij}_n \cos (n (\theta_{a,i}- \epsilon \theta_{b,j})) + \delta^{ij} \ln \left\lvert 2 \sin \left(\frac{\theta_{a,i}- \epsilon \theta_{b,i}}{2}\right)\right\rvert \right] \\
     & \left. + \sum_{i\in\ver} \sum_{a=1}^{N_i}\gamma^{i}_{\text{measure}} (\theta_{a,i}) \right\} .
\end{aligned}
\end{equation}
The last line collects the contribution to the Haar measure that depend on a single eigenvalue, plus, for $G_i \in \{SO(2N_i), SO(2N_i+1)\}$, a term which removes the undue summand $(\theta_{b,i}=\theta_{a,i}, \epsilon=-1)$ in the Vandermonde factor in the second line. Explicitly:
\begin{equation}\label{eq:gammameasureshift}
    \gamma^{i}_{\text{measure}} (\theta) = \begin{cases} -\ln \left( 2 \sin \theta\right) & G_i = SO(2N_i) \\ \ln \left( 2 \sin \frac{\theta}{2}\right)-\ln \left( 2 \sin \theta\right)& G_i = SO(2N_i+1) \\  \ln \left( 2 \sin \theta\right) & G_i = USp(2N_i) . \end{cases}
\end{equation}
In the orthosymplectic case, $\theta \in [0,\pi]$, thus the eigenvalue densities \eqref{eq:defrhoi} satisfy the normalisation condition
\begin{equation}\label{eq:normalisationSOSprho}
    \int_0^{\pi} \rho_i (\theta) \dd \theta = \frac{N_i}{N} .
\end{equation}
It is convenient to use a change of variables of the form $x=\cos \theta$ and expand in the basis of Chebyshev polynomials $T_n (x)$, which satisfy the orthogonality relations 
\begin{equation}
    \int_{-1}^{1} \frac{\dd x}{\sqrt{1-x^2}} T_n (x) T_m (x) = \frac{\pi}{2} \delta_{mn} \qquad n \geq 1 .
\end{equation}
We then introduce $\varrho_i (x)$, such that $\rho_i (\theta)=\varrho_i (\cos \theta)$, and expand it in Chebyshev polynomials:
\begin{equation}
    \varrho_i (x) = \frac{N_i}{\pi N} + \sum_{n\geq1} \chi_{i,n} T_n (x) ,
\end{equation}
where the $0^{\text{th}}$ order is fixed by \eqref{eq:normalisationSOSprho}. After manipulations analogous to the unitary matrix models in section \ref{sec:General} and using elementary trigonometric identities, we arrive at:
\begin{equation}\label{eq:SeffFourierSOSp}
\begin{aligned}
    S_{\mathrm{eff}}^{\mathfrak{osp}} = \pi^2 N^2 \sum_{n\ge1}\frac{1}{n}\sum_{i \in \ver} &\left[ \frac{\chi_{i,n}}{\pi N} \left( v^{i}_n - \sigma_n^{i} + \sum_{j \in \ver} \kappa_j h^{ij}_n \right) +\frac{1}{2}  \sum_{j\in \ver} \chi_{i,n}\chi_{j,n} \left( h^{ij}_n - \delta^{ij} \right)\right] ,
\end{aligned}
\end{equation}
where the shift $\sigma_n^{i}$, defined in the main text in \eqref{defshiftSOSp}, originates from the mode expansion of \eqref{eq:gammameasureshift}.
The saddle point equations stemming from \eqref{eq:SeffFourierSOSp} are
\begin{equation}\label{eq:SOSpSPE}
    \frac{1}{\pi N}\left( v^{i}_n - \sigma_n^{i} + \sum_{i^{\prime} \in \ver} \kappa_{i^{\prime}}  h^{ii^{\prime}}_n \right) +  \sum_{j\in \ver} \chi_{j,n} \left( \frac{h^{ij}_n+h^{ji}_n}{2} - \delta^{ij} \right) =0 ,
\end{equation}
$\forall i \in \ver, n \in \N$. In our intended application to theories with 8 supercharges, the coefficients are symmetric, $h^{ij}_n=h^{ji}_n$. Proceeding in complete analogy with section \ref{sec:FiniteL} from this point on, we define 
\begin{equation}
    D_n ^{\mathfrak{osp}} = \left[ \mathbb{I} - h_n \right]^{-1} .
\end{equation}
It is also convenient to define the shorthand notation 
\begin{equation}
    \tilde{v}^{i}_n = v^{i}_n - \sigma_n^{i} + \sum_{i^{\prime} \in \ver} \kappa_{i^{\prime}}  h^{ii^{\prime}}_n  .
\end{equation}
Notice that $\tilde{v}^{i}_n$ contains all the effective single-trace contributions to the integrand: $v^{i}_n$ is a coefficient of the single-trace contributions to \eqref{eq:fSOSp}, $\sigma_n^{j}$ comes from the measure, and the last sum comes from collecting the contributions of the trivial eigenvalues of $SO(2N_{i^{\prime}}+1)$ to the double-trace terms. In applications, one can split $v^{i}_n=v^{i}_{\Box,n}+v^{i}_{\circlearrowleft,n}$, where the first contribution comes from the framing of the quiver and the second from the collection of all contributions to the double-trace terms which effectively depend only on one eigenvalue.\par
With these definitions, \eqref{eq:SOSpSPE} is solved by 
\begin{equation}
    \chi_{i,n} = \frac{1}{\pi N} \sum_{j\in \ver} \left[D_n^{\mathfrak{osp}} \right]_{ij} \tilde{v}^{j}_n ,
\end{equation}
and the effective action evaluated on the saddle point configuration gives
\begin{equation}\label{eq:SeffSOSpSPE}
    \left.S_{\mathrm{eff}}^{\mathfrak{osp}} \right\rvert_{\text{saddle}}= \frac{1}{2}\sum_{n\ge1}\frac{1}{n}\sum_{(i,j) \in \ver^2}  \left[D_n ^{\mathfrak{osp}} \right]_{ij} \tilde{v}^{i}_n\tilde{v}^{j}_n .
\end{equation}
Integrating out the Gaussian fluctuations of the modes $\chi_{i,n}$ around the saddle point configuration (notice that there are half of them, compared to the unitary case), we finally arrive at 
\begin{equation}\label{eq:ZSOSpuniversal}
    \left.\ln \mI \right\rvert_{N\to\infty} = \frac{1}{2}\sum_{n=1}^{\infty} \left( \frac{1}{n} \sum_{i,j \in \ver} \left[ D_n^{\mathfrak{osp}} \right]_{ij} \tilde{v}^{i}_n \tilde{v}^{j}_n  +\ln \det_{i,j \in \ver}\left[ D_n^{\mathfrak{osp}} \right]_{ij}  + \frac{2}{n}c_n\right) ,
\end{equation}
which proves the result \eqref{eq:ISOSpuniversal} stated in the main text. 

\section{Balanced long quivers near \texorpdfstring{$t \to 1$}{t=1}}
\label{app:balancedt}
In section \ref{sec:longbalanced} we are interested in the limit $t\to 1$ of $\PE\left[ g (t) \right]$, where 
\begin{equation}
    g(t) = -\left(LN^2\int \dd z\dd z'\nu(z)\varphi(z')G_1(z,z')\right)(1-t)^2 .
\end{equation}\par
We set $t=e^{-\alpha}$ and assume throughout that  
\begin{equation}\label{eq:talphacondition}
    \Re (\alpha ) > 0 , \qquad - \pi < \Im \left(\alpha\right) <\pi , 
\end{equation}
which guarantees that $\Re(M_1)\ne 0$ in \eqref{eq:Greenmass}.
It is easy to show, using the explicit dependence of $G_1(z,z')$ on $t$ through \eqref{eq:Greenmass}, that there exist $0<\delta_0,\delta_1 <1$ and constants $C_0,C_1$ such that
\begin{equation}
\begin{aligned}
    \text{(i) } &\lvert g(t)\rvert < C_0 \lvert t \rvert, \qquad \qquad \forall 0 < \lvert t\rvert< \delta_0 \\
    \text{(ii) }&\lvert g(t)\rvert < C_1 \lvert 1- t\rvert^2, \qquad \forall 0 < \lvert 1-t\rvert < \delta_1 \\
    \text{(iii) }&g(e^{-\alpha}) \text{ is a smooth function of $\alpha$ in \eqref{eq:talphacondition}}. 
\end{aligned}
\end{equation}
(i) follows from the explicit solution \eqref{eq:Gnlinear} for the Green's function, which for $\Re(M_1) \ne 0$ behaves near $t\to 0$ as 
\begin{equation}
    (1-t)^2G_1 (z,z') \underset{t \to 0}{\sim} \frac{L}{2} \sqrt{t} \exp \left( -\frac{L}{ \lvert\Re (\sqrt{t})\rvert} \lvert z-z'\rvert \right) ,
\end{equation}
thus the integrand in $g(t)$ is exponentially suppressed away from $z-z'$. (ii) follows immediately from \eqref{eq:Gnlinear}, which near $t\to 1$ yields $g(t) = C_2 (1-t)^2 + \mathcal{O}\left((1-t)^4\right)$.\par
The limit $t \to 1$ is taken by sending $\lvert\alpha\rvert \to 0^+$ from within the sector \eqref{eq:talphacondition}. We find
\begin{equation}
    \lim_{\lvert \alpha\rvert \to 0^+} \PE \left[ g(t) \right] =  \lim_{\rvert\alpha\rvert\to 0^+} \exp \left( \sum_{n \geq 1} \frac{g (e^{-n\alpha)}}{n} \right)  = \exp \left( \int \frac{\dd r}{r} g (e^{-r}) \right) .
\end{equation}
The integration contour goes from 0 to $\infty$ in the complex plane, with $- \pi < \Im \left(r\right) <\pi $. The integrand is convergent by the above properties: (i) guarantees that the integrand falls off fast enough at $r \to \infty$, (ii) guarantees the integrability at $r \to 0$, and (iii) along the contour satisfying \eqref{eq:talphacondition} the integrand is smooth.\par
We have thus shown that $\PE \left[ g(t) \right]$ has a finite $t\to 1$ limit, at least when approached from $t$ such that $\Re (M_1) \ne 0$, establishing the claim in section \ref{sec:longbalanced}.

\end{appendix}
\bibliography{ref}

@article{Santilli:2025zum,
    author = "Santilli, Leonardo",
    title = "{Large $N$ Limits of Supersymmetric Quantum Field Theories: A Pedagogical Overview}",
    eprint = "2501.05794",
    archivePrefix = "arXiv",
    primaryClass = "hep-th",
    doi = "10.1002/prop.70006",
    journal = "Fortsch. Phys.",
    volume = "73",
    number = "6",
    pages = "e70006",
    year = "2025"
}

@article{friends,
    author = "Fang, Yuanyuan and Feng, Jing and Xie, Dan",
    title = "{Seiberg dualities for quiver gauge theories}",
    eprint = "2609.*****",
    archivePrefix = "arXiv",
    primaryClass = "hep-th",
    year = "2026"
}

@article{Reid-Edwards:2010vpm,
    author = "Reid-Edwards, R. A. and Stefanski, jr., B.",
    title = "{On Type IIA geometries dual to $N = 2$ SCFTs}",
    eprint = "1011.0216",
    archivePrefix = "arXiv",
    primaryClass = "hep-th",
    doi = "10.1016/j.nuclphysb.2011.04.002",
    journal = "Nucl. Phys. B",
    volume = "849",
    pages = "549--572",
    year = "2011"
}

@article{Coccia:2021lpp,
    author = "Coccia, Lorenzo and Uhlemann, Christoph F.",
    title = "{Mapping out the internal space in AdS/BCFT with Wilson loops}",
    eprint = "2112.14648",
    archivePrefix = "arXiv",
    primaryClass = "hep-th",
    doi = "10.1007/JHEP03(2022)127",
    journal = "JHEP",
    volume = "03",
    pages = "127",
    year = "2022"
}

@article{Benvenuti:2010pq,
    author = "Benvenuti, Sergio and Hanany, Amihay and Mekareeya, Noppadol",
    title = "{The Hilbert Series of the One Instanton Moduli Space}",
    eprint = "1005.3026",
    archivePrefix = "arXiv",
    primaryClass = "hep-th",
    doi = "10.1007/JHEP06(2010)100",
    journal = "JHEP",
    volume = "06",
    pages = "100",
    year = "2010"
}

@article{Gray:2008yu,
    author = "Gray, James and Hanany, Amihay and He, Yang-Hui and Jejjala, Vishnu and Mekareeya, Noppadol",
    title = "{SQCD: A Geometric Apercu}",
    eprint = "0803.4257",
    archivePrefix = "arXiv",
    primaryClass = "hep-th",
    reportNumber = "IHES-P-08-04, IMPERIAL-TP-08-AH-03",
    doi = "10.1088/1126-6708/2008/05/099",
    journal = "JHEP",
    volume = "05",
    pages = "099",
    year = "2008"
}

@article{tHooft:1973alw,
    author = "'t Hooft, Gerard",
    editor = "Taylor, J. C.",
    title = "{A Planar Diagram Theory for Strong Interactions}",
    reportNumber = "CERN-TH-1786",
    doi = "10.1016/0550-3213(74)90154-0",
    journal = "Nucl. Phys. B",
    volume = "72",
    pages = "461",
    year = "1974"
}

@article{Romelsberger:2005eg,
    author = "Romelsberger, Christian",
    title = "{Counting chiral primaries in N = 1, d=4 superconformal field theories}",
    eprint = "hep-th/0510060",
    archivePrefix = "arXiv",
    doi = "10.1016/j.nuclphysb.2006.03.037",
    journal = "Nucl. Phys. B",
    volume = "747",
    pages = "329--353",
    year = "2006"
}

@article{Kinney:2005ej,
    author = "Kinney, Justin and Maldacena, Juan Martin and Minwalla, Shiraz and Raju, Suvrat",
    title = "{An Index for 4 dimensional super conformal theories}",
    eprint = "hep-th/0510251",
    archivePrefix = "arXiv",
    doi = "10.1007/s00220-007-0258-7",
    journal = "Commun. Math. Phys.",
    volume = "275",
    pages = "209--254",
    year = "2007"
}

@article{Bhattacharya:2008zy,
    author = "Bhattacharya, Jyotirmoy and Bhattacharyya, Sayantani and Minwalla, Shiraz and Raju, Suvrat",
    title = "{Indices for Superconformal Field Theories in 3,5 and 6 Dimensions}",
    eprint = "0801.1435",
    archivePrefix = "arXiv",
    primaryClass = "hep-th",
    reportNumber = "TIFR-TH-08-01, HUTP-08-A0001",
    doi = "10.1088/1126-6708/2008/02/064",
    journal = "JHEP",
    volume = "02",
    pages = "064",
    year = "2008"
}

@article{Gadde:2020yah,
    author = "Gadde, Abhijit",
    title = "{Lectures on the Superconformal Index}",
    eprint = "2006.13630",
    archivePrefix = "arXiv",
    primaryClass = "hep-th",
    reportNumber = "TIFR/TH/20-20",
    doi = "10.1088/1751-8121/ac42ac",
    journal = "J. Phys. A",
    volume = "55",
    number = "6",
    pages = "063001",
    year = "2022"
}

@article{Gadde:2010en,
    author = "Gadde, Abhijit and Rastelli, Leonardo and Razamat, Shlomo S. and Yan, Wenbin",
    title = "{On the Superconformal Index of $N=1$ IR Fixed Points: A Holographic Check}",
    eprint = "1011.5278",
    archivePrefix = "arXiv",
    primaryClass = "hep-th",
    reportNumber = "YITP-SB-10-38",
    doi = "10.1007/JHEP03(2011)041",
    journal = "JHEP",
    volume = "03",
    pages = "041",
    year = "2011"
}

@article{Bergman:2013koa,
    author = "Bergman, Oren and Rodr\'\i{}guez-G\'omez, Diego and Zafrir, Gabi",
    title = "{5d superconformal indices at large $N$ and holography}",
    eprint = "1305.6870",
    archivePrefix = "arXiv",
    primaryClass = "hep-th",
    doi = "10.1007/JHEP08(2013)081",
    journal = "JHEP",
    volume = "08",
    pages = "081",
    year = "2013"
}

@article{Dolan:2008qi,
    author = "Dolan, F. A. and Osborn, H.",
    title = "{Applications of the Superconformal Index for Protected Operators and q-Hypergeometric Identities to $N=1$ Dual Theories}",
    eprint = "0801.4947",
    archivePrefix = "arXiv",
    primaryClass = "hep-th",
    reportNumber = "DAMTP-08-07, DIAS-STP-08-02, SHEP-08-06",
    doi = "10.1016/j.nuclphysb.2009.01.028",
    journal = "Nucl. Phys. B",
    volume = "818",
    pages = "137--178",
    year = "2009"
}

@article{Kim:2012gu,
    author = "Kim, Hee-Cheol and Kim, Sung-Soo and Lee, Kimyeong",
    title = "{5-dim Superconformal Index with Enhanced En Global Symmetry}",
    eprint = "1206.6781",
    archivePrefix = "arXiv",
    primaryClass = "hep-th",
    reportNumber = "KIAS-P12033",
    doi = "10.1007/JHEP10(2012)142",
    journal = "JHEP",
    volume = "10",
    pages = "142",
    year = "2012"
}

@article{Eager:2012hx,
    author = "Eager, Richard and Schmude, Johannes and Tachikawa, Yuji",
    title = "{Superconformal Indices, Sasaki-Einstein Manifolds, and Cyclic Homologies}",
    eprint = "1207.0573",
    archivePrefix = "arXiv",
    primaryClass = "hep-th",
    reportNumber = "IPMU-12-0135, UT-12-18",
    doi = "10.4310/ATMP.2014.v18.n1.a3",
    journal = "Adv. Theor. Math. Phys.",
    volume = "18",
    number = "1",
    pages = "129--175",
    year = "2014"
}

@article{Brezin:1977sv,
    author = "Brezin, E. and Itzykson, C. and Parisi, G. and Zuber, J. B.",
    title = "{Planar Diagrams}",
    reportNumber = "SACLAY-DPH-T-77-126",
    doi = "10.1007/BF01614153",
    journal = "Commun. Math. Phys.",
    volume = "59",
    pages = "35",
    year = "1978"
}

@article{tHooft:1974pnl,
    author = "'t Hooft, Gerard",
    title = "{A Two-Dimensional Model for Mesons}",
    reportNumber = "CERN-TH-1820",
    doi = "10.1016/0550-3213(74)90088-1",
    journal = "Nucl. Phys. B",
    volume = "75",
    pages = "461--470",
    year = "1974"
}

@article{Veneziano:1976wm,
    author = "Veneziano, G.",
    title = "{Some Aspects of a Unified Approach to Gauge, Dual and Gribov Theories}",
    reportNumber = "CERN-TH-2200",
    doi = "10.1016/0550-3213(76)90412-0",
    journal = "Nucl. Phys. B",
    volume = "117",
    pages = "519--545",
    year = "1976"
}

@article{Beccaria:2023qnu,
    author = "Beccaria, M. and Korchemsky, G. P.",
    title = "{Four-dimensional $\mathcal{N}$ = 2 superconformal long circular quivers}",
    eprint = "2312.03836",
    archivePrefix = "arXiv",
    primaryClass = "hep-th",
    reportNumber = "IPhT-T23/113",
    doi = "10.1007/JHEP04(2024)054",
    journal = "JHEP",
    volume = "04",
    pages = "054",
    year = "2024"
}

@article{Korchemsky:2025eyc,
    author = "Korchemsky, Gregory P. and Testa, Alessandro",
    title = "{Correlation functions in four-dimensional superconformal long circular quivers}",
    eprint = "2501.17223",
    archivePrefix = "arXiv",
    primaryClass = "hep-th",
    doi = "10.1007/JHEP07(2025)223",
    journal = "JHEP",
    volume = "07",
    pages = "223",
    year = "2025"
}

@article{Hayling:2017cva,
    author = "Hayling, Joseph and Papageorgakis, Constantinos and Pomoni, Elli and Rodr{\'\i}guez-G{\'o}mez, Diego",
    title = "{Exact Deconstruction of the 6D (2,0) Theory}",
    eprint = "1704.02986",
    archivePrefix = "arXiv",
    primaryClass = "hep-th",
    reportNumber = "QMUL-PH-17-06, DESY-17-030",
    doi = "10.1007/JHEP06(2017)072",
    journal = "JHEP",
    volume = "06",
    pages = "072",
    year = "2017"
}

@article{Gadde:2011uv,
    author = "Gadde, Abhijit and Rastelli, Leonardo and Razamat, Shlomo S. and Yan, Wenbin",
    title = "{Gauge Theories and Macdonald Polynomials}",
    eprint = "1110.3740",
    archivePrefix = "arXiv",
    primaryClass = "hep-th",
    reportNumber = "YITP-SB-11-30",
    doi = "10.1007/s00220-012-1607-8",
    journal = "Commun. Math. Phys.",
    volume = "319",
    pages = "147--193",
    year = "2013"
}

@article{Hong:2018amk,
    author = "Hong, Junho and Liu, James T. and Mayerson, Daniel R.",
    title = "{Gauged Six-Dimensional Supergravity from Warped IIB Reductions}",
    eprint = "1808.04301",
    archivePrefix = "arXiv",
    primaryClass = "hep-th",
    doi = "10.1007/JHEP09(2018)140",
    journal = "JHEP",
    volume = "09",
    pages = "140",
    year = "2018"
}

@article{Nunez:2019gbg,
    author = "N{\'u}{\~n}ez, Carlos and Roychowdhury, Dibakar and Speziali, Stefano and Zacar{\'\i}as, Salom{\'o}n",
    title = "{Holographic aspects of four dimensional ${\cal N }=2$ SCFTs and their marginal deformations}",
    eprint = "1901.02888",
    archivePrefix = "arXiv",
    primaryClass = "hep-th",
    doi = "10.1016/j.nuclphysb.2019.114617",
    journal = "Nucl. Phys. B",
    volume = "943",
    pages = "114617",
    year = "2019"
}

@article{Gaiotto:2009gz,
    author = "Gaiotto, Davide and Maldacena, Juan",
    title = "{The Gravity duals of N=2 superconformal field theories}",
    eprint = "0904.4466",
    archivePrefix = "arXiv",
    primaryClass = "hep-th",
    doi = "10.1007/JHEP10(2012)189",
    journal = "JHEP",
    volume = "10",
    pages = "189",
    year = "2012"
}

@article{Aharony:2012tz,
    author = "Aharony, Ofer and Berdichevsky, Leon and Berkooz, Micha",
    title = "{4d N=2 superconformal linear quivers with type IIA duals}",
    eprint = "1206.5916",
    archivePrefix = "arXiv",
    primaryClass = "hep-th",
    reportNumber = "WIS-11-12-JUNE-DPPA",
    doi = "10.1007/JHEP08(2012)131",
    journal = "JHEP",
    volume = "08",
    pages = "131",
    year = "2012"
}

@article{Legramandi:2021uds,
    author = "Legramandi, Andrea and Nunez, Carlos",
    title = "{Electrostatic description of five-dimensional SCFTs}",
    eprint = "2104.11240",
    archivePrefix = "arXiv",
    primaryClass = "hep-th",
    doi = "10.1016/j.nuclphysb.2021.115630",
    journal = "Nucl. Phys. B",
    volume = "974",
    pages = "115630",
    year = "2022"
}

@article{Akhond:2021ffz,
    author = "Akhond, Mohammad and Legramandi, Andrea and Nunez, Carlos",
    title = "{Electrostatic description of 3d $ \mathcal{N} $ = 4 linear quivers}",
    eprint = "2109.06193",
    archivePrefix = "arXiv",
    primaryClass = "hep-th",
    doi = "10.1007/JHEP11(2021)205",
    journal = "JHEP",
    volume = "11",
    pages = "205",
    year = "2021"
}

@article{Legramandi:2021aqv,
    author = "Legramandi, Andrea and Nunez, Carlos",
    title = "{Holographic description of SCFT$_{5}$ compactifications}",
    eprint = "2109.11554",
    archivePrefix = "arXiv",
    primaryClass = "hep-th",
    doi = "10.1007/JHEP02(2022)010",
    journal = "JHEP",
    volume = "02",
    pages = "010",
    year = "2022"
}

@article{DHoker:2016ujz,
    author = "D'Hoker, Eric and Gutperle, Michael and Karch, Andreas and Uhlemann, Christoph F.",
    title = "{Warped $AdS_6\times S^2$ in Type IIB supergravity I: Local solutions}",
    eprint = "1606.01254",
    archivePrefix = "arXiv",
    primaryClass = "hep-th",
    doi = "10.1007/JHEP08(2016)046",
    journal = "JHEP",
    volume = "08",
    pages = "046",
    year = "2016"
}

@article{Uhlemann:2019ypp,
    author = "Uhlemann, Christoph F.",
    title = "{Exact results for 5d SCFTs of long quiver type}",
    eprint = "1909.01369",
    archivePrefix = "arXiv",
    primaryClass = "hep-th",
    doi = "10.1007/JHEP11(2019)072",
    journal = "JHEP",
    volume = "11",
    pages = "072",
    year = "2019"
}

@article{Uhlemann:2020bek,
    author = "Uhlemann, Christoph F.",
    title = "{Wilson loops in 5d long quiver gauge theories}",
    eprint = "2006.01142",
    archivePrefix = "arXiv",
    primaryClass = "hep-th",
    reportNumber = "LCTP-20-10",
    doi = "10.1007/JHEP09(2020)145",
    journal = "JHEP",
    volume = "09",
    pages = "145",
    year = "2020"
}

@article{Coccia:2020wtk,
    author = "Coccia, Lorenzo and Uhlemann, Christoph F.",
    title = "{On the planar limit of 3d $ {\mathrm{T}}_{\rho}^{\sigma}\left[\mathrm{SU}\left(\mathrm{N}\right)\right] $}",
    eprint = "2011.10050",
    archivePrefix = "arXiv",
    primaryClass = "hep-th",
    doi = "10.1007/JHEP06(2021)038",
    journal = "JHEP",
    volume = "06",
    pages = "038",
    year = "2021"
}

@article{DHoker:2017mds,
    author = "D'Hoker, Eric and Gutperle, Michael and Uhlemann, Christoph F.",
    archivePrefix = "arXiv",
    doi = "10.1007/JHEP05(2017)131",
    eprint = "1703.08186",
    journal = "JHEP",
    pages = "131",
    primaryClass = "hep-th",
    title = "{Warped $AdS_6\times S^2$ in Type IIB supergravity II: Global solutions and five-brane webs}",
    volume = "05",
    year = "2017"
}

@article{DHoker:2017zwj,
    author = "D'Hoker, Eric and Gutperle, Michael and Uhlemann, Christoph F.",
    archivePrefix = "arXiv",
    doi = "10.1007/JHEP11(2017)200",
    eprint = "1706.00433",
    journal = "JHEP",
    pages = "200",
    primaryClass = "hep-th",
    title = "{Warped $AdS_6\times S^2$ in Type IIB supergravity III: Global solutions with seven-branes}",
    volume = "11",
    year = "2017"
}

@article{Uhlemann:2019lge,
    author = "Uhlemann, Christoph F.",
    title = "{AdS$_6$/CFT$_5$ with O7-planes}",
    eprint = "1912.09716",
    archivePrefix = "arXiv",
    primaryClass = "hep-th",
    reportNumber = "LCTP-19-34",
    doi = "10.1007/JHEP04(2020)113",
    journal = "JHEP",
    volume = "04",
    pages = "113",
    year = "2020"
}

@article{He:2024djr,
    author = "He, Dongming and Uhlemann, Christoph F.",
    title = "{Solving $ \mathcal{N} $ = 4 SYM BCFT matrix models at large N}",
    eprint = "2409.13016",
    archivePrefix = "arXiv",
    primaryClass = "hep-th",
    doi = "10.1007/JHEP12(2024)164",
    journal = "JHEP",
    volume = "12",
    pages = "164",
    year = "2024"
}

@article{Akhond:2022awd,
    author = "Akhond, Mohammad and Legramandi, Andrea and Nunez, Carlos and Santilli, Leonardo and Schepers, Lucas",
    title = "{Massive flows in AdS6/CFT5}",
    eprint = "2211.09824",
    archivePrefix = "arXiv",
    primaryClass = "hep-th",
    doi = "10.1016/j.physletb.2023.137899",
    journal = "Phys. Lett. B",
    volume = "840",
    pages = "137899",
    year = "2023"
}

@article{Akhond:2022oaf,
    author = "Akhond, Mohammad and Legramandi, Andrea and Nunez, Carlos and Santilli, Leonardo and Schepers, Lucas",
    title = "{Matrix Models and Holography: Mass Deformations of Long Quiver Theories in 5d and 3d}",
    eprint = "2211.13240",
    archivePrefix = "arXiv",
    primaryClass = "hep-th",
    doi = "10.21468/SciPostPhys.15.3.086",
    journal = "SciPost Phys.",
    volume = "15",
    pages = "086",
    year = "2023"
}

@article{Santilli:2023fuh,
    author = "Santilli, Leonardo and Uhlemann, Christoph F.",
    title = "{3d defects in 5d: RG flows and defect F-maximization}",
    eprint = "2305.01004",
    archivePrefix = "arXiv",
    primaryClass = "hep-th",
    doi = "10.1007/JHEP06(2023)136",
    journal = "JHEP",
    volume = "06",
    pages = "136",
    year = "2023"
}

@article{Nunez:2023loo,
    author = "Nunez, Carlos and Santilli, Leonardo and Zarembo, Konstantin",
    title = "{Linear Quivers at Large-$N$}",
    eprint = "2311.00024",
    archivePrefix = "arXiv",
    primaryClass = "hep-th",
    doi = "10.1007/s00220-024-05186-1",
    journal = "Commun. Math. Phys.",
    volume = "406",
    number = "1",
    pages = "6",
    year = "2025"
}

@article{Fatemiabhari:2022kpv,
    author = "Fatemiabhari, Ali and Nunez, Carlos",
    title = "{Wilson loops for 5d and 3d conformal linear quivers}",
    eprint = "2209.07536",
    archivePrefix = "arXiv",
    primaryClass = "hep-th",
    doi = "10.1016/j.nuclphysb.2023.116125",
    journal = "Nucl. Phys. B",
    volume = "989",
    pages = "116125",
    year = "2023"
}

@article{Apruzzi:2022nax,
    author = "Apruzzi, Fabio and Bergman, Oren and Kim, Hee-Cheol and Uhlemann, Christoph F.",
    title = "{Generalized quotients and holographic duals for 5d S-fold SCFTs}",
    eprint = "2211.13243",
    archivePrefix = "arXiv",
    primaryClass = "hep-th",
    reportNumber = "LCTP-22-15",
    doi = "10.1007/JHEP04(2023)027",
    journal = "JHEP",
    volume = "04",
    pages = "027",
    year = "2023"
}

@article{Santilli:2020ueh,
    author = "Santilli, Leonardo and Tierz, Miguel",
    title = "{Exact equivalences and phase discrepancies between random matrix ensembles}",
    eprint = "2003.10475",
    archivePrefix = "arXiv",
    primaryClass = "math-ph",
    doi = "10.1088/1742-5468/aba594",
    journal = "J. Stat. Mech.",
    volume = "2008",
    pages = "083107",
    year = "2020"
}

@article{Santilli:2021eon,
    author = "Santilli, Leonardo and Tierz, Miguel",
    title = "{Multiple phases and meromorphic deformations of unitary matrix models}",
    eprint = "2102.11305",
    archivePrefix = "arXiv",
    primaryClass = "hep-th",
    doi = "10.1016/j.nuclphysb.2022.115694",
    journal = "Nucl. Phys. B",
    volume = "976",
    pages = "115694",
    year = "2022"
}

@article{Calderon-Infante:2026rkj,
    author = "Calder{\'o}n-Infante, Jos{\'e} and Mohseni, Amineh",
    title = "{The CFT distance conjecture and tensionless string limits in $ \mathcal{N} $ = 2 quiver gauge theories}",
    eprint = "2601.08909",
    archivePrefix = "arXiv",
    primaryClass = "hep-th",
    doi = "10.1007/JHEP04(2026)105",
    journal = "JHEP",
    volume = "04",
    pages = "105",
    year = "2026"
}

@article{Rastelli:2016tbz,
    author = "Rastelli, Leonardo and Razamat, Shlomo S.",
    title = "{The supersymmetric index in four dimensions}",
    eprint = "1608.02965",
    archivePrefix = "arXiv",
    primaryClass = "hep-th",
    doi = "10.1088/1751-8121/aa76a6",
    journal = "J. Phys. A",
    volume = "50",
    number = "44",
    pages = "443013",
    year = "2017"
}

@article{Amariti:2020jyx,
    author = "Amariti, Antonio and Fazzi, Marco and Segati, Alessia",
    title = "{The SCI of $ \mathcal{N} $ = 4 USp(2N$_{c}$) and SO(N$_{c}$) SYM as a matrix integral}",
    eprint = "2012.15208",
    archivePrefix = "arXiv",
    primaryClass = "hep-th",
    doi = "10.1007/JHEP06(2021)132",
    journal = "JHEP",
    volume = "06",
    pages = "132",
    year = "2021"
}

@article{Amariti:2021ubd,
    author = "Amariti, Antonio and Fazzi, Marco and Segati, Alessia",
    title = "{Expanding on the Cardy-like limit of the SCI of 4d $ \mathcal{N} $ = 1 ABCD SCFTs}",
    eprint = "2103.15853",
    archivePrefix = "arXiv",
    primaryClass = "hep-th",
    doi = "10.1007/JHEP07(2021)141",
    journal = "JHEP",
    volume = "07",
    pages = "141",
    year = "2021"
}

@book{ForresterBook,
    AUTHOR = {Forrester, Peter J.},
     TITLE = {Log-gases and random matrices},
    SERIES = {London Mathematical Society Monographs Series},
    VOLUME = {34},
 PUBLISHER = {Princeton University Press, Princeton, NJ},
      YEAR = {2010},
     PAGES = {xiv+791},
       DOI = {10.1515/9781400835416},
}

@book{CourantHilbertBook,
    AUTHOR = {Courant, R. and Hilbert, D.},
     TITLE = {Methods of Mathematical Physics},
    VOLUME = {1},
 PUBLISHER = {Wiley},
      YEAR = {1924},
     PAGES = {xv+560},
       DOI = {10.1002/9783527617210},
}

@article{Bergman:2012kr,
    author = "Bergman, Oren and Rodriguez-Gomez, Diego",
    title = "{5d quivers and their AdS(6) duals}",
    eprint = "1206.3503",
    archivePrefix = "arXiv",
    primaryClass = "hep-th",
    doi = "10.1007/JHEP07(2012)171",
    journal = "JHEP",
    volume = "07",
    pages = "171",
    year = "2012"
}

@article{GonzalezLezcano:2026wje,
    author = "Gonz{\'a}lez Lezcano, Alfredo and Pando Zayas, Leopoldo A. and Ray, Augniva",
    title = "{The giant graviton expansion in AdS$_5\times$SE$_5$}",
    eprint = "2607.02658",
    archivePrefix = "arXiv",
    primaryClass = "hep-th",
    month = "7",
    year = "2026"
}

@article{Benini:2006hh,
    author = "Benini, Francesco and Canoura, Felipe and Cremonesi, Stefano and Nunez, Carlos and Ramallo, Alfonso V.",
    title = "{Unquenched flavors in the Klebanov-Witten model}",
    eprint = "hep-th/0612118",
    archivePrefix = "arXiv",
    reportNumber = "SISSA-78-2006-EP, US-FT-5-06",
    doi = "10.1088/1126-6708/2007/02/090",
    journal = "JHEP",
    volume = "02",
    pages = "090",
    year = "2007"
}

@article{Benini:2007gx,
    author = "Benini, Francesco and Canoura, Felipe and Cremonesi, Stefano and Nunez, Carlos and Ramallo, Alfonso V.",
    title = "{Backreacting flavors in the Klebanov-Strassler background}",
    eprint = "0706.1238",
    archivePrefix = "arXiv",
    primaryClass = "hep-th",
    reportNumber = "SISSA-36-2007-EP, US-FT-3-07",
    doi = "10.1088/1126-6708/2007/09/109",
    journal = "JHEP",
    volume = "09",
    pages = "109",
    year = "2007"
}

@article{Kapustin:1998xn,
    author = "Kapustin, Anton",
    title = "{Solution of N=2 gauge theories via compactification to three-dimensions}",
    eprint = "hep-th/9804069",
    archivePrefix = "arXiv",
    reportNumber = "IASSNS-HEP-98-34",
    doi = "10.1016/S0550-3213(98)00520-3",
    journal = "Nucl. Phys. B",
    volume = "534",
    pages = "531--545",
    year = "1998"
}

@article{Benini:2010uu,
    author = "Benini, Francesco and Tachikawa, Yuji and Xie, Dan",
    title = "{Mirrors of 3d Sicilian theories}",
    eprint = "1007.0992",
    archivePrefix = "arXiv",
    primaryClass = "hep-th",
    reportNumber = "MIFPA-10-27, PUTP-2344",
    doi = "10.1007/JHEP09(2010)063",
    journal = "JHEP",
    volume = "09",
    pages = "063",
    year = "2010"
}

@article{Xie:2012hs,
    author = "Xie, Dan",
    title = "{General Argyres-Douglas Theory}",
    eprint = "1204.2270",
    archivePrefix = "arXiv",
    primaryClass = "hep-th",
    doi = "10.1007/JHEP01(2013)100",
    journal = "JHEP",
    volume = "01",
    pages = "100",
    year = "2013"
}

@article{Kang:2022zsl,
    author = "Kang, Monica Jinwoo and Lawrie, Craig and Lee, Ki-Hong and Sacchi, Matteo and Song, Jaewon",
    title = "{Higgs branch, Coulomb branch, and Hall-Littlewood index}",
    eprint = "2207.05764",
    archivePrefix = "arXiv",
    primaryClass = "hep-th",
    reportNumber = "CALT-TH-2022-024, DESY-22-110, CALT-TH-2022-024; DESY-22-110",
    doi = "10.1103/PhysRevD.106.106021",
    journal = "Phys. Rev. D",
    volume = "106",
    number = "10",
    pages = "106021",
    year = "2022"
}

@article{Beem:2017ooy,
    author = "Beem, Christopher and Rastelli, Leonardo",
    title = "{Vertex operator algebras, Higgs branches, and modular differential equations}",
    eprint = "1707.07679",
    archivePrefix = "arXiv",
    primaryClass = "hep-th",
    reportNumber = "YITP-SB-17-27",
    doi = "10.1007/JHEP08(2018)114",
    journal = "JHEP",
    volume = "08",
    pages = "114",
    year = "2018"
}

@article{Arkani-Hamed:2001wsh,
    author = "Arkani-Hamed, Nima and Cohen, Andrew G. and Kaplan, David B. and Karch, Andreas and Motl, Lubos",
    title = "{Deconstructing (2,0) and little string theories}",
    eprint = "hep-th/0110146",
    archivePrefix = "arXiv",
    reportNumber = "HUTP-01-A049",
    doi = "10.1088/1126-6708/2003/01/083",
    journal = "JHEP",
    volume = "01",
    pages = "083",
    year = "2003"
}

@article{Purkayastha:2025whi,
    author = "Purkayastha, Souradeep and Qu, Zishen and Zahabi, Ali",
    title = "{Quiver superconformal index and giant gravitons: asymptotics and expansions}",
    eprint = "2509.12123",
    archivePrefix = "arXiv",
    primaryClass = "hep-th",
    doi = "10.1007/JHEP07(2026)070",
    journal = "JHEP",
    volume = "07",
    pages = "070",
    year = "2026"
}

@article{Bourget:2020xdz,
    author = "Bourget, Antoine and Grimminger, Julius F. and Hanany, Amihay and Kalveks, Rudolph and Sperling, Marcus and Zhong, Zhenghao",
    title = "{Magnetic Lattices for Orthosymplectic Quivers}",
    eprint = "2007.04667",
    archivePrefix = "arXiv",
    primaryClass = "hep-th",
    reportNumber = "Imperial/TP/20/AH/07",
    doi = "10.1007/JHEP12(2020)092",
    journal = "JHEP",
    volume = "12",
    pages = "092",
    year = "2020"
}
\end{document}